\documentclass[journal]{IEEEtran}
\pdfoutput=1
\usepackage{amsmath,amsfonts}
\usepackage{textcomp}
\usepackage{stfloats}
\usepackage{url}
\usepackage{verbatim}
\usepackage{cite}
\usepackage{amssymb}
\usepackage{algorithm,eqparbox,array}
\usepackage[T1]{fontenc}
\usepackage{graphicx}
\usepackage{graphbox}
\usepackage[noend]{algpseudocode}
\usepackage{array}
\usepackage{mathtools}
\usepackage{xcolor}
\usepackage{stmaryrd}
\usepackage{moreverb}
\usepackage{lineno}
\usepackage{tabularx}
\graphicspath{ {images/} }
\usepackage{multirow}
\usepackage[utf8]{inputenc}
\usepackage{stackengine}
\usepackage{float}
\usepackage{epstopdf}
\usepackage{makecell}
\usepackage[colorlinks]{hyperref}
\usepackage{lipsum}
\usepackage{setspace}
\usepackage[english]{babel}
\usepackage{microtype}
\usepackage{subfig}
\makeatletter
\renewcommand{\fnum@figure}{Fig. \thefigure}
\makeatother

\begin{document}

\title{A Deep Reinforcement Learning based Algorithm for Time and Cost Optimized Scaling of Serverless Applications}

\author{Anupama Mampage,
        Shanika Karunasekera,
        and~Rajkumar~Buyya
\thanks{The authors are with the Cloud Computing and Distributed Systems (CLOUDS) Laboratory, School of Computing and Information Systems, The University of Melbourne, Australia.\protect\\
E-mail: mampage@student.unimelb.edu.au, \\
karus@unimelb.edu.au, 
rbuyya@unimelb.edu.au}}



\maketitle

\begin{abstract}
Serverless computing has gained a strong traction in the cloud computing community in recent years. Among the many benefits of this novel computing model, the rapid auto-scaling capability of user applications takes prominence. However, the offer of adhoc scaling of user deployments at function level introduces many complications to serverless systems. The added delay and failures in function request executions caused by the time consumed for dynamically creating new resources to suit function workloads, known as the cold-start delay, is one such very prevalent shortcoming.  Maintaining idle resource pools to alleviate this issue often results in wasted resources from the cloud provider perspective. Existing solutions to address this limitation mostly focus on predicting and understanding function load levels in order to proactively create required resources. Although these solutions improve function performance, the lack of understanding on the overall system characteristics in making these scaling decisions often leads to the sub-optimal usage of system resources. Further, the multi-tenant nature of serverless systems requires a scalable solution adaptable for multiple co-existing applications, a limitation seen in most current solutions. In this paper, we introduce a novel multi-agent Deep Reinforcement Learning based intelligent solution for both horizontal and vertical scaling of function resources, based on a comprehensive understanding on both function and system requirements. Our solution elevates function performance reducing cold starts, while also offering the flexibility for optimizing resource maintenance cost to the service providers. Experiments conducted considering varying workload scenarios show improvements of up to 23\% and 34\% in terms of application latency and request failures, while also saving up to 45\% in infrastructure cost for the service providers.
\end{abstract}


\section{Introduction}
\IEEEPARstart{S}{erverless} computing has been embraced as an application deployment model favorable for many application domains in the current world \cite{mampage2022holistic}. The provider centric resource management  model has succeeded in attaining the "serverless" nature of operations for the end user. However, the cloud provider is tasked with numerous added responsibilities as never before in achieving this seemingly "serverless" behavior of cloud systems. Rapid auto-scalability of user applications in line with load variations, is among the highly valued distinguishing properties of a serverless computing platform, which has proven useful under many application scenarios. The very fine-grained auto-scaling capabilities in serverless platforms require deployed functions to scale their resources just-in-time, as user demand varies. As such, function resources would scale to zero, when there is no request traffic and scale back up when needed, ensuring high resource efficiency. Setting up new resources in this manner on the go, results in a considerable start up time, widely known as the problem of the 'cold start delay' in functions which hinders its performance. Cold start delay in essence, is a combination of the function runtime environment set up time and the time spent on application specific code initialization. This initial delay becomes specially significant for serverless functions with very low execution times, which is the majority. The situation is further complicated by the existence of multiple user applications deployed on the same infrastructure, which require individual attention in their scaling decisions.

A number of existing works have studied the auto-scaling techniques employed by both the commercial and open source serverless computing platforms, and how they affect application performance \cite{manner2018cold}, \cite{vahidinia2020cold}. \cite{wang2018peeking} compares AWS Lambda, Google Cloud Functions and Microsoft Azure in terms of their function cold start delay. These platforms maintain idle function instances from previous executions for a  particular time duration before recycling, in order to have more ready-to-serve warm instances for new executions. AWS and Google seem to have relatively stable cold start delays while Azure platform showed more varying values at the time of their experimentation. Relationships also exist between factors such as the programming language used and the memory size of a function instance and the resulting resource start up delays. The majority of open source serverless frameworks including Fission \cite{Fission47:online}, Kubeless \cite{Kubeless28:online}, OpenFaas \cite{HomeOpen89:online} and Knative \cite{HomeKnat2:online} are built utilizing Kubernetes \cite{Kubernet0:online} as the function orchestrator \cite{mohanty2018evaluation}. The auto-scaling functionality of these frameworks is usually based on a set resource utilization threshold of the existing function instances or the number of requests per second, which determines the required number of function replicas required to meet the current load.

Research works which address the issues related to serverless auto-scaling delays are identified under two categories. One set of solutions is directed towards reducing the frequency of the occurrence of cold start delays, while the other is focused on reducing the measured cold start delay of an individual function instance \cite{vahidinia2020cold}. In order to reduce the delay itself, various techniques are presented to improve sandbox creation times, including the creation of the required network elements beforehand, utilizing snapshots of previously used containers and designing and developing customized sandbox environments \cite{mohan2019agile}, \cite{silva2020prebaking}, \cite{oakes2018sock}. On the other hand, minimizing the frequency of cold starts is achieved by employing techniques for creating pre-warmed containers, reusing warm containers and adjusting the level of concurrently served requests by a function instance. These approaches often times try to predict the arrival rates and demand levels for individual functions in order to proactively create the required resources \cite{stein2018serverless}, \cite{singhvi2021atoll}, \cite{ling2019pigeon}. The techniques used for such predictions mostly incorporate the resource consumption characteristics of the serverless functions in order to determine the size of the resource pool to be maintained. They rarely consider the resource availability status or the cost of maintaining such idle resource pools to the serverless resource provider. The distinguished billing model in serverless platforms favours its end users by charging them only when the resources are actively being used with a millisecond level accuracy. This means that even though additional resource pools are maintained to meet Quality of Service (QoS) requirements of the user, the provider is able to recover the costs of such resources only to the extent of them being used, calculated at a very fine level. As such, careful calculations are required considering the status of the platform resources along with function characteristics, in order to make these scaling decisions. Moreover, the majority of solutions are limited in their capability of handling multi-tenancy in the function scaling process.

Considering the above highlighted challenges, our work is focused on carrying out the scaling of function resources of multiple user applications in a way that would enhance application performance, while at the same time, preserving the optimum usage of cloud provider resources. Further, while existing solutions are designed to support only horizontal scaling of function instances,  i.e., scaling in or out the number of function replicas, our solution approach encapsulates both horizontal and vertical scaling, for better optimizing our target objectives. Vertical scaling handles scaling up and down of the cpu and memory capacities of the function resources. In addition to varying the number of function instances to meet changing user request rates, adapting the resource configuration of existing function instances to handle the incoming traffic in this manner helps in balancing our dual objectives of function performance and provider cost optimization.

Deep Reinforcement Learning (DRL) techniques are being extensively explored for cloud resource management work from recent times. Experience based learning encouraged in the RL paradigm makes it a good candidate as a method of learning the behavior of dynamic serverless workloads with very short execution durations. In this work, we propose a DRL based solution which employs multiple learning agents to determine the optimum level of function scaling to suit changing demand levels. The key \textbf{contributions} of our work are as follows:
\begin{enumerate}
\item We formulate and present a RL based model of the function auto-scaling problem in a multi-tenant serverless computing environment.

\item We propose a novel multi-agent function scaling framework based on the policy gradient algorithm Asynchronous Advantage Actor Critic (A3C), which aims to attain a balance in optimizing application performance and provider resource cost. We adapt the A3C algorithm to suit a multi-discrete action space required in making the horizontal and vertical scaling decisions for a multitude of user applications residing in the platform at a time.

\item We train and evaluate our DRL model in a python based simulator environment. We also design a practical testbed based on the open-source serverless platform Kubeless which is deployed on a Kubernetes cluster. The simulator replicates the characteristics and behavior of the practical testbed and utilizes function profiling data derived from the same, in all its experiments.

\item We evaluate and compare our approach with baseline scaling techniques using real world serverless applications, together with function traces captured from Microsoft Azure Functions. 

\end{enumerate}
The rest of the paper is organized as follows: Section II reviews relevant literature. Section III presents the system model and formulates the function scaling problem mathematically. Section IV introduces the proposed DRL oriented scaling framework. Sections V and VI discuss the design and implementation details of the DRL agent training environment, evaluation of the proposed technique and the scope for future work.

\section{Related Work}

\begingroup
\begin{normalsize}
\setlength{\tabcolsep}{10pt} 
\renewcommand{\arraystretch}{1.3}
\begin{table*}
	\caption{Summary of Literature Review}
	\label{table:relatedWork}
	\resizebox{\textwidth}{!}{\begin{tabular}{c c c c c c c c c c c c}
			\hline
			\\
			\multicolumn{1}{c}{\textbf{Work}}
			&\multicolumn{2}{c}{\textbf{\centering Application Model}}
			&\multicolumn{1}{c}{\textbf{\centering Scaling}}
                &\multicolumn{2}{c}{\textbf{\centering Scaling Type}}
			&\multicolumn{4}{c}{\textbf{Decision Parameters}}
			&\multicolumn{1}{c}{\textbf{Multi}}
			&\multicolumn{1}{c}{\textbf{VM }}
			\\
			\cline{2-3} \cline{5-6} \cline{7-10}
			& Single & Function& Technique& Horizontal & Vertical
			& \multicolumn{2}{c}{Optimization Objective} & Workload &Overall System 
			& Tenancy &Heterogeneity \\
			
			\cline{7-8}
			& Function &Chain &&&& Response Time & Provider Cost Efficiency & Awareness & Awareness \\
			\hline
			\cite{stein2018serverless} &\checkmark& & Heuristic &\checkmark & &\checkmark &\checkmark &\checkmark&&\checkmark  \\
                \cite{ling2019pigeon} &\checkmark&& Heuristic &\checkmark &&\checkmark&&&&\\
                \cite{lin2019mitigating} &\checkmark&&Heuristic&\checkmark&&\checkmark\\
                \cite{xu2019adaptive} &&\checkmark&ML&\checkmark&&\checkmark&\checkmark&\checkmark&&\\
			\cite{bermbach2020using} &&\checkmark&Heuristic&\checkmark&&\checkmark&&\checkmark&&&\checkmark\\
                \cite{solaiman2020wlec} &\checkmark&& Heuristic &\checkmark &&\checkmark&&\checkmark&&\checkmark\\
                \cite{daw2020xanadu} &&\checkmark& Heuristic &\checkmark &&\checkmark&\checkmark&\checkmark&&\\
                 \cite{somma2020less} &\checkmark&&Q-Learning&\checkmark&&\checkmark\\
                \cite{agarwal2021reinforcement} &\checkmark&&Q-Learning&\checkmark&&\checkmark\\
                \cite{singhvi2021atoll} &\checkmark& \checkmark &Mathematical modelling&\checkmark&&\checkmark&&&&\checkmark\\
                \cite{suo2021tackling} &\checkmark&&Mathematical modelling&\checkmark&&\checkmark&&\checkmark&&\checkmark\\
                \cite{yu2021harvesting} &\checkmark&&PPO&&\checkmark&\checkmark&&\checkmark&&\checkmark\\
                \cite{schuler2021ai} &\checkmark&&Q-Learning&&\checkmark&\checkmark&\\
                \cite{li2022kneescale} &\checkmark&&Heuristic&\checkmark&&\checkmark&&\checkmark&&\checkmark&\\
                \cite{phung2022prediction} &\checkmark&&Mathematical Modelling&\checkmark&&\checkmark&&\checkmark&\\
                \cite{zafeiropoulos2022reinforcement} &\checkmark&&Q-Learning/DQN&\checkmark&&\checkmark\\
                \cite{benedetti2022reinforcement} &\checkmark&&Q-Learning&\checkmark&&\checkmark&&\checkmark&&&\checkmark\\
                \cite{vahidinia2022mitigating} &\checkmark&&A2C/Mathematical Modelling&\checkmark&&\checkmark&&\checkmark\\
                \cite{qiu2022reinforcement} &\checkmark&&MA-PPO&\checkmark&\checkmark&\checkmark&&\checkmark&&\checkmark\\
                \cite{zhang2022adaptive} &\checkmark&&Q-Learning/Heuristic&\checkmark&\checkmark&\checkmark&&\checkmark&\\
			Our proposed work &  \checkmark& \checkmark  & MA-A3C & \checkmark&\checkmark&\checkmark&\checkmark & \checkmark&\checkmark&\checkmark&\checkmark\\
			\hline
	\end{tabular}}
\end{table*}
\end{normalsize}
\endgroup

\subsection{Serverless Resource Scaling} 
Scaling of serverless functions could be discussed in terms of the horizontal and vertical scaling aspects. Horizontal scaling refers to varying the number of instances of a particular function that is available for request execution. As demand levels vary for an application with time, determining the optimum level of replica scaling required to meet the target objectives is a challenging task. \cite{stein2018serverless} try to predict the required number of function instances in order to keep the new request waiting time below a set threshold, by using a heuristic technique. \cite{singhvi2021atoll} use an exponentially weighted moving average model to estimate request arrival rates. Proactive allocation of sandboxes is done using this estimate. An oversubscribed static resource pool with pre-warmed containers of all resource sizes is proposed in \cite{ling2019pigeon}. \cite{bermbach2020using} implement a lightweight middleware which uses the knowledge of function compositions to trigger cold starts, leading to provisioning of new containers before they are required. A container management system with three queues containing cold and warm containers based on their features is introduced in \cite{solaiman2020wlec}. \cite{daw2020xanadu}, \cite{lin2019mitigating}, \cite{suo2021tackling} and \cite{xu2019adaptive} propose maintaining a pool of function instances to face request demands. A heuristic solution is given in \cite{li2022kneescale} to adjust the replica number without compromising on user budget.  Time-series forecasting is used in \cite{phung2022prediction} to determine the request workload to support the scaling decisions. 

Q-Learning based approaches are used in \cite{somma2020less}, \cite{agarwal2021reinforcement} and \cite{zafeiropoulos2022reinforcement} to determine the number of function containers to scale-up/down at each point in time in order to maintain low application latency and failure rates. \cite{benedetti2022reinforcement} use Q-Learning to decide the optimum level of maximum cpu usage in a function instance to trigger scaling. In \cite{vahidinia2022mitigating} the DRL algorithm A2C is used to determine the idle time window for a used function instance and further a time series model is used to predict future invocations and thereby, create warm containers.

Vertical scaling deals with the up/down scale of the resource capacities of a function instance. This is seen as an alternative or used in conjunction with horizontal scaling in order to meet intended targets, in the face of changing traffic levels. An actor critic architecture with Proximal Policy Optimization (PPO) is used in \cite{yu2021harvesting} to harvest idle resources from functions and direct them to under-provisioned instances. A Q-Learning based solution is given in \cite{schuler2021ai} to identify the level of concurrency, i.e the number of concurrent requests served per instance, to optimize function latency and system throughput. A DRL based multi-agent (MA) solution is analysed against a single agent implementation in \cite{qiu2022reinforcement} for the horizontal and vertical scaling of functions. They focus on function latency and resource efficiency for users and thereby lack focus on the overall platform resource utilization. A preliminary study is done in \cite{zhang2022adaptive} on using Q-Learning for horizontal scaling decisions and a heuristic approach for vertical scaling. 


\subsection{RL Solutions for Serverless Resource Management}

As also discussed above in section II(A), a number of recent works employ RL techniques for enhancing resource management in serverless environments. The target areas of improvement in this manner include resource scheduling, scaling and modelling of optimum resource configurations for functions.

A policy gradient algorithm is proposed in \cite{yu2021faasrank}, to identify the best node for scheduling a function request. \cite{dehury2021def} uses a DRL approach to determine the percentage of user requests to be processed by the cloud and offloaded to the fog layer. A distributed task scheduling approach is presented in \cite{tang2022distributed} for serverless edge computing networks. They explore a multi-agent dueling Deep Q Learning (DQN) architecture to assist the edge network in making resource allocation and scheduling decisions. A distributed, experience-sharing, function offloading framework for the edge is proposed in \cite{yao2023performance}. They suggest an improved actor critic algorithm for deciding whether to execute functions on the IoT device or on an edge device. \cite{jeon2021deep} introduce a multi-agent DRL solution for caching packages required for running serverless functions at edge nodes, based on their importance and popularity. They aim to improve per function response time while managing resources consumed while caching. A multi-step DQN based solution is proposed in \cite{mampage2023deep} for function scheduling, in a multi-tenant serverless environment, which aims to optimize application performance as well as provider resource cost.

We summarize the reviewed works specifically in the area of serverless resource scaling in Table \ref{table:relatedWork}. This comparison considers the aspects of application model, used technique, type of scaling, optimization objective, workload-awareness (consideration for request arrival rate fluctuations), system awareness (knowledge on individual cluster VM resource usage metrics), multi-tenancy (adaptability to suit multiple concurrent applications) and VM heterogeneity. Majority of existing works propose solutions based on only one aspect of scaling, either horizontal or vertical, which is not ideal for optimizing resource cost. Thus such solutions lack the motivation to be used in their serverless platforms by service providers. Further, many solutions lack overall system status awareness, and also are not scalable for decision making under a multi-tenant scenario. In contrast, our work captures the dynamic function workload as well as system parameters, with adaptability to suit multi-tenant clusters, and targets optimizing dual objectives concerning both the user and the provider.

\section{Adaptive Function Scaling}

\subsection{System Model}

\begin{figure}[!b]
	\centering 
	\includegraphics[width=\linewidth, height=7.5cm]{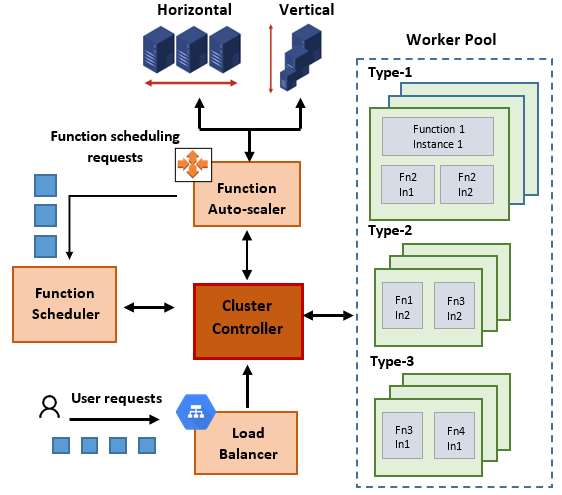}
	\caption{The system model of the serverless application execution environment}
	\label{fig:SystemModel}
\end{figure}

Our system model represents the common serverless system architecture in use across a majority of open-source serverless frameworks \cite{HomeOpen89:online}, \cite{HomeKnat2:online}, \cite{Fission47:online}. The main components of this architecture include, a cluster of Virtual Machines (VMs), a load balancer acting as the entry point for user requests, a function auto-scaler, a function scheduler and a controller which coordinates the communication between all the other functional units. The highlevel system model is illustrated in Fig. \ref{fig:SystemModel}. 

We consider our worker pool to be composed of a set of heterogeneous VMs with varying compute and memory capacities. An instance of a function is deployed in a container and managed as a single resource unit called a pod. A pod is able to serve multiple concurrent function requests depending on its resource capacity. Creation of replicas of the same function type and adjusting the resource configuration of an existing instance is triggered by the auto-scalar depending on the implemented technique of scaling, which is the focus of this work. Subsequently, the function scheduler selects a suitable node and schedules the created instance. 

User requests enter the system via the load balancer, which queues them and directs them to existing function replicas. In the absence of suitable resources, the requests are dropped after the passage of a certain time duration after arrival. The load balancer is considered to be distributing the incoming requests among the existing replicas in a round-robin manner. Serverless applications are formed of either a single or multiple functions. Multi function applications are composed of chained functions which are executed sequentially. Thus, an instance of a function may serve requests of multiple user applications. Requests are received at the deployed functions in a stochastic manner, and thus the demand levels for a function instance could vary rapidly in a very short time. Performance degradation caused by cold start delays are imminent, if the auto-scaling mechanism is not capable of pro-actively determining the scale up of resources required on time. At the same time, gross over-estimation of resources and over-provisioning of the same, lead to massive inefficiencies in resource maintenance cost for the provider. Thus at a given time, the auto-scaler needs to decide the ideal number of replicas and the resource configuration of each replica of a particular function, that would help reach a satisfactory balance in function performance and provider cost.

\subsection{Problem Formulation}

Suppose $N$ is the total number of VMs in a serverless cluster. Each VM is of varying size in terms of its CPU (number of vCPU cores) and memory (MB) capacity. $Q$ is the total number of different function types deployed in the cluster. Let $P^{k}$ and $Req^{k}$ denote a pod/instance and a single request of the $k^{th}$ ($k \in [1, Q]$) function respectively, while $M^{k}(t)$ is the number of existing pods of the same at time $t$. Each function instance of type $k$ has four attributes at time $t$, i.e., allocated pod CPU ($P^{k}_{cpu}(t)$), pod memory ($P^{k}_{mem}(t)$), CPU and memory consumption of a single request ($Req^{k}_{cpu}$ and $Req^{k}_{mem}$), standard response time ($r^{k}_{0}$) and the arrival rate of requests of the type. The standard response time for a request refers to the average request response time for a function when executed in a pre-created pod without any resource creation delays. 

Compute power is often identified to be a main source of resource pressure in serverless functions, leading to poor application performance \cite{suresh2020ensure}. Based on this logic, the default horizontal auto-scaler in our system model triggers a new pod creation and scaling down of existing pods based on a target average CPU utilization of a pod of that type, $T^{k}_{cpu.util}(t)$. i.e., if the number of new pods of type $k$ to be created is $N^{k}_{\Delta}$ and the maximum allowed number of pods of any type at any time is $M^{max}$, 

\begin{equation}
\label{eq:1}
N^{k}_{\Delta}(t) = min [M^{k}(t) \times \frac{C^{k}_{cpu.util}(t)}{T^{k}_{cpu.util}(t)}, M^{max}] - M^{k}(t)
\end{equation}

where $C^{k}_{cpu.util}(t)$ refers to the current average pod CPU utilization. Our task in terms of horizontal scaling is to determine the ideal $T^{k}_{cpu.util}(t)$ value at a given time. Pods which are not currently being used are scaled down as required where $N^{k}_{\Delta}(t)$ is a negative value.

Along with the action of the horizontal scaler, the vertical auto-scaler needs to determine the best suited levels of CPU and memory configurations for a function of type $k$ at a given time $t$. The incremental/decremental CPU value $cpu^{k}_{\Delta}(t)$ and the memory value $mem^{k}_{\Delta}(t)$ which form the vertical scaling decision, need to meet a few constraints, i.e, 

(1) the resulting resource allocation levels after action execution need to be within the upper and lower boundaries of applicable CPU and memory resource limits to a function instance, 

\begin{equation}
\label{eq:2}
\begin{split}
P_{cpu.min} < P^{k}_{cpu}(t) + cpu^{k}_{\Delta}(t) < P_{cpu.max}  \\
 &\  \hspace{-65mm} P_{mem.min} < P^{k}_{mem}(t) + mem^{k}_{\Delta}(t) < P_{mem.max}
 \end{split}
\end{equation}

 We consider these allocated resources for a function instance to be hard limits, i.e., these mark upper limits of resource consumption by a single pod, irrespective of the traffic levels.

(2) The chosen resource increments need to be compatible with the available resource levels of VMs holding the existing function replicas, 

\begin{equation}
\label{eq:3}
[P^{k}_{cpu}(t) + cpu^{k}_{\Delta}(t)] \times M^{k}_{vm^{i}}(t) < {vm_{cpu}^{i}}(t)   \hspace{0.5cm}   \forall i \in [1, N]
\end{equation}

\begin{equation}
\label{eq:4}
[P^{k}_{mem}(t) + mem^{k}_{\Delta}(t)] \times M^{k}_{vm^{i}}(t) < {vm_{mem}^{i}}(t)  \hspace{0.5cm}    \forall i \in [1, N]
\end{equation}

where $M^{k}_{vm^{i}}(t)$ is the number of replicas of $k^{th}$ function residing in $vm^{i}$, while ${vm_{cpu}^{i}}(t)$ and ${vm_{mem}^{i}}(t)$ is the available CPU and memory of the same at time $t$.

(3) The resource configuration change in an instance should not affect the function requests already in execution. Thus the new resource allocation should not go below the current resource utilization levels of any pod of the type.

\begin{equation}
\label{eq:5}
 P^{kj}_{cpu.util}(t) < [P^{k}_{cpu}(t) + cpu^{k}_{\Delta}(t)]  \hspace{0.5cm} \forall j \in [1, M^{k}(t)]
\end{equation}

\begin{equation}
\label{eq:6}
 P^{kj}_{mem.util}(t) < [P^{k}_{mem}(t) + cpu^{k}_{\Delta}(t)]   \hspace{0.5cm}  \forall j \in [1, M^{k}(t)]
\end{equation}

where $P^{kj}_{cpu.util}$ and $P^{kj}_{mem.util}$ are the cpu and memory utilization levels of the $j^{th}$ pod of function $k$. The time $t$ in the above expressions: (\ref{eq:1}), (\ref{eq:2}), (\ref{eq:3}), (\ref{eq:4}), (\ref{eq:5}), and (\ref{eq:6}) indicates the time steps in which scaling decisions are taken.

\begingroup
\setlength{\tabcolsep}{2pt} 
\renewcommand{\arraystretch}{1} 
\begin{table}[!t]
\caption{Definition of Symbols}
\label{table:symbols}
\footnotesize
\begin{tabular*}{\columnwidth}{l|l}
\hline
\textbf{Symbol} & \textbf{Definition} \\
\hline
\textit{$N$} & {Total number of available VMs} \\
\textit{$Q$} & {Total number of deployed functions} \\
\textit{$P_{cpu}^{k}$} & {Allocated CPU for a pod of the $k^{th}$ function, $k \in [1, Q]$} \\
\textit{$P_{mem}^{k}$} & {Allocated memory for a pod of the $k^{th}$ function, $k \in [1, Q]$} \\
\textit{$M^{k}_{vm_{i}}$} & {Number of existing pods of the $k^{th}$ function residing in $vm_{i}$} \\
\textit{$M^{max}$} & {Maximum allowed number of pods of any single function type} \\
\textit{$P^{kj}_{cpu.util}$} & {Cpu utilization of the $j^{th}$ pod of function $k$, $j \in [1, M^{k}]$} \\
\textit{$P^{kj}_{mem.util}$} & {Memory utilization of the $j^{th}$ pod of function $k$, $j \in [1, M^{k}]$} \\
\textit{$T^{k}_{cpu.util}$} & {Target average CPU utilization of a pod of type $k$} \\
\textit{$C^{k}_{cpu.util}$} & {Current average CPU utilization of a pod of type $k$} \\
\textit{$N^{k}_{\Delta}$} & {Number of new pods of type $k$ to be created} \\
\textit{$cpu^{k}_{\Delta}$} & {Change in allocated CPU to a pod of type $k$} \\
\textit{$mem^{k}_{\Delta}$} & {Change in allocated memory to a pod of type $k$} \\
\textit{$vm_{cpu}^{i}$} & {Available cpu in $vm_{i}$ } \\
\textit{$vm_{mem}^{i}$} & {Available memory in $vm_{i}$ } \\
\textit{$A$} & {Total number of user applications deployed} \\
\textit{$V^{b}$} & {Number of user requests received by application $b$, $b \in [1, A]$} \\
\textit{$R^{b}_{q}$} & {The sum of response times of each function relevant to the}\\& {${q}^{th}$ request of an application, $q \in [1, V^{b}]$} \\
\textit{$R^{b}_{q0}$} & {Total standard response time of the constituent functions} \\&{of the ${q}^{th}$ request of an application} \\
\textit{$price_{i}$} & {Unit price of VM, $v_i$} \\
\textit{$t_{i}$} & {Total active time of VM, $v_i$} \\

\hline
\end{tabular*}
\label{tab1}
\end{table}
\endgroup

One target objective of this work is to minimize sub-optimal application performance caused by the lack of a proper resource scaling strategy. As mentioned previously, the resources allocated to function instances are set as hard limits, which prevents them from causing resource contention in the host node, with increased traffic levels. Thus, we could consider that the performance degradation of applications under such a system model is a direct effect of the absence of enough ready resources to face request demand levels. Cold start delays introduced by new resource creation affect request response times and may also cause request failures. Hence we consider application response time latency and request failure rates to be metrics which directly reflect the effects of the platform scaling decisions. 

Let A be the total number of user applications deployed in the platform. Consider $V^b$, $1 \leq b \leq A$ to be the total request traffic to $b^{th}$ application. The sum of response times of the constituent functions corresponding to the ${q}^{th}$ request of application $b$ ($1 \leq q \leq V^{b}$) is $R^{b}_{q}$. Also, the estimated total standard response time of the same is denoted by $R^{b}_{q0}$. The ratio of $R^{b}_{q}$ and $R^{b}_{q0}$ averaged over the total number of requests received by an application, is called the average Relative Application Response Time (RART). We aim to minimize the average RART, calculated across all the deployed applications over the duration of a workload. Here we define RART instead of the response time itself, to eliminate any bias in our optimization objective arising from execution time variations in serverless functions.

\begin{equation}
\label{eq:7}
\begin{split}
Average\hspace{1mm} RART  =  {\frac{1}{A}\sum_{b=1}^{A}\frac{1}{V^b}{\sum_{q=1}^{V^b} {\frac{R^{b}_{q}}{R^{b}_{q0}}}}} 
\end{split}
\end{equation}

The sum of response times of each function which form the considered execution sequence of an application, is used for calculating $R^{b}_{q}$ and $R^{b}_{q0}$. The preceding function in a chained application simply evokes the next function in the sequence. Hence this calculation is possible as this process is devoid of any data communication delay. The total standard response time for an application's request (${R^{b}_{q0}}$) is expressed as a function of $q$, since the relevant function sequence is dependent on the request input. 

Further, as part of performance optimization, we aim to minimize Request Failure Rates (RFR), i.e., the number of dropped function requests as a ratio of the total requests received. Accordingly we express our performance optimization objective as follows:

\begin{equation}
\label{eq:8}
\begin{split}
Minimize: [Average\hspace{1mm} RART, RFR]
\end{split}
\end{equation}



In this work we also plan to optimize the infrastructure cost of the provider. In the calculation of the resource costs we incorporate VM instance pricing, as we consider a heterogeneous serverless cluster composed of VMs of different CPU and memory sizes. The provider cost optimization objective is formulated as follows:

\begin{equation}
\label{eq:9}
\begin{split}
Minimize: Cost_{Total} = {\sum_{i=1}^{N} price_{i} \times t_{i}}\\
\end{split}
\end{equation}

$price_{i}$ is the unit price of VM $v_i$ and $t_{i}$ is the total time that the $i^{th}$ VM was active during workload executions. A VM is considered to be active, when it is serving requests of at least one function. Thus resource efficiency is achieved when active VMs have high utilization levels.

Our primary objectives of minimizing function performance degradation and enabling high resource efficiency tend to be conflicting objectives. Therefore, we utilize a system parameter $\beta \in [0, 1]$, so that users can achieve a sufficient trade-off between the two. Accordingly, our overall target objective is as follows:

\begin{equation}
\label{eq:10}
\begin{split}
    Minimize: \beta \times Sum\hspace{1mm} (Average\hspace{1mm} RART + RFR)+\\
    &\ \hspace{-27mm}(1-\beta) \times Cost_{Total}
\end{split}
\end{equation}

Table \ref{table:symbols} summarizes the various symbols introduced in this section.  

\section{Reinforcement Learning Model}

\subsection{Learning Model for Function Scaling}

RL is a branch of machine learning that encourages an experience based learning style. A RL agent interacts with its environment and at each time step takes an action $a_{t}$, based on the current policy $\pi(a_{t}|s_{t})$, where $s_{t}$ is the current state of the environment. A reward $r_{t+1}$ is received in turn based on the 'goodness' of the action. The RL agent's final objective is to learn a policy which maximizes the cumulative reward over a sequence of actions. 

In this work we explore the applicability of the concept of RL in developing an adaptive scaling policy for applications in a multi-tenant serverless computing environment. The RL agent takes the role of forming the basis of scaling each function, either horizontally or vertically, with variations in user demand levels. The serverless platform forms the environment with which the agent communicates and derives state information at each time step. Time steps in which these scaling configuration changes are executed, are considered to happen at regular intervals. The received reward after each scaling action implementation is dependent on the target level of optimization of each of the dual objectives discussed above. The key aspects of the RL model in the context of our problem are discussed below.

\textbf{State space}: The state information needs to encapsulate both the resource metrics of the serverless platform infrastructure as well as the resource requirements, traffic levels and the current performance of the function to be scaled. Accordingly, the state in our environment could be presented as a 1-dimensional vector, where the first part describes the cluster VM specifications: [$vm^{i}_{cpu.util}$, $vm^{i}_{mem.util}$, $vm^{i}_{cpu.alloc}$, $vm^{i}_{mem.alloc}$, $vm^{i}_{cpu.cap}$, $vm^{i}_{mem.cap}$, $vm^{i}_{replicas}$ ]. $vm^{i}_{cpu.util}$ and $vm^{i}_{mem.util}$ refer to the actual cpu and memory utilization levels of the VMs, $vm^{i}_{cpu.alloc}$ and $vm^{i}_{mem.alloc}$ refer to the percentage of cpu and memory that is allocated to pods, $vm^{i}_{cpu.cap}$ and $vm^{i}_{mem.cap}$ denote the cpu and memory capacities (this is representative of the VM unit prices), while $vm^{i}_{replicas}$ represent the number of replicas of the scaling function, that is currently present in the VM. These VM resource metrics are gathered for all the cluster VMs to form the state space. Note that the resource utilization and allocation levels identify as two separate metrics since although resources are allocated to function pods, the VM resources actually utilized depend on the function requests in execution in those pods. The second part of the state vector is composed of the function specifications: [$P^{k}_{cpu}$, $P^{k}_{mem}$, $Req^{k}_{cpu}$, $Req^{k}_{mem}$, $P^{k}_{rate}$,  $P^{k}_{RFRT}$, $P^{k}_{RFR}$, $C^{k}_{cpu.util}$, $C^{k}_{mem.util}$]. $P^{k}_{cpu}$ and $P^{k}_{mem}$ represent the requested cpu and memory by a funciton instance, $Req^{k}_{cpu}$ and $Req^{k}_{mem}$ represent the resource consumption of a single request of the type, $P^{k}_{rate}$ represents the current request rate, $P^{k}_{RFRT}$ represent the Relative Function Response Time (RFRT), which is the ratio of the actual function response time to the standard response time, $P^{k}_{RFR}$ represent the function request failure rate, while $C^{k}_{cpu.util}$ and $C^{k}_{mem.util}$ represent the average cpu and memory utilization of all the pods of type $k$. These metrics when consolidated, give the RL agent a comprehensive understanding on the current system status, in order to reach the best scaling policy with time. At the start of each time step, the agent gathers the required data and forms this state vector before determining the scaling action. 

\textbf{Action space}: We model the action space in our environment as a novel multi-discrete action space where we need to determine three decision parameters namely, the target average cpu utilization value for triggering horizontal function scaling ($a_{1}$), the change in allocated cpu ($a_{2}$), and memory ($a_{3}$) values for pods of the considered function type. Since combining the three actions to formulate an action space with all possible combinations leads to an explosion in the action space size, we consider the three to be independent decision variables. Further we discretize each variable to suit the scale of our modelled environment, where each action would reflect either an increase, decrease or maintaining the same level in the particular variable. Accordingly, a complete action generated by the DRL agent could be presented as [$a_{1}$, $a_{2}$, $a_{3}$].

\textbf{Reward}: The reward assigned to the agent at each step immediately after an action, needs to resonate with the target objectives of optimization. Since the considered objectives of performance and provider cost optimization in this work usually compete with each other and thus could be conflicting, we define two separate reward structures for the two. Accordingly, the reward for action $a_{t}$ is:

1. \textbf{$R_{1}$}: The sum of the average RFRT and RFR of all the deployed functions in the cluster a set time interval after the implementation of action $a_{t}$. Since application response time is a function of that of its constituent functions, RFRT acts as a proxy to measure our performance optimization objective of RART in Eq.(\ref{eq:7}). In addition, it is more closely identifiable with each function scaling decision of the DRL agent.

2. \textbf{$R_{2}$}: The difference in the total cluster VM up time cost (Eq.(\ref{eq:9})) just before and a set time interval after the implementation of action $a_{t}$.

Since the cumulative of both these reward values at the end of an episode needs to be minimized in order to reach our target improvements, we insert a negative sign to motivate reduction in latency and cost over time. Further, at each step we normalize the three values of RFRT, RFR, and VM cost which are in different scales in order to remove any notion of being biased towards one value, in the process of DRL model training. We derive the minimum and maximum values for each of these step rewards after running and observing these values over many workload scenarios. As such, the awarded reward to the agent after each scaling decision is as follows:

\begin{equation}
\label{eq:11}
     Reward = -( ( \beta \times R_{1normalized}) + ((1-\beta) \times R_{2normalized}))
\end{equation}

\subsection{Actor-Critic based Multi-agent Scaling Framework}

\begin{algorithm}[tb]
	\caption{Actor-Critic based Multi-agent Scaling Algorithm} 
	\label{algo:A3Calgorithm}
	\begin{algorithmic}[1]
	    \State{Initialize the global shared actor and critic network parameters $\theta$ and $\phi$}            
            \For{worker = 1 to N}
	    \State{Initialize the local actor and critic network parameters $\theta'$ and $\phi'$}
            \State{Initialize the local step counter t = 0}
            \State{Initialize the training parameters $\alpha$, $\gamma$ and network update frequency $f$}
            \State {Initialize the local training environment for the worker agent}
            \For{episode = 1 to E}
	    \State{Reset the environment}
	    \For{step = 1 to T}
            \State Input the state $s$ of the environment to actor network $\pi_{\theta'}(a|s)$
            \For{i = 1 to 3}
            \State Select action $a_i$ using the marginal distribution $\pi_{{\theta'}_{i}}(a|s)$
            \EndFor
            \State Execute the combined action ($a = a_1, a_2, a_3$), move to the next state $s'$ and observe the reward $r$
            \State Store the transition ($s,a,r,s'$) in memory $D$
            \If{$t  \%  f == 0$ or step = T} 
            \For{j = 1 to K}
            \State Compute the advantage estimates $A^{\hat{}}_1$ to $A^{\hat{}}_K$
            \State Compute the loss and the gradients of the loss of actor $\nabla_{\theta'}J(\theta')$ and critic $\nabla_{\phi'}J(\phi')$ networks
            \EndFor
            \State Perform asynchronous update of global actor and critic network parameters $\theta$ and $\phi$
            \State Synchronize the local actor and critic network parameters $\theta'$ and $\phi'$ with $\theta$ and $\phi$
            \State Clear memory D
            \EndIf
            \State $t \leftarrow t + 1$
            \EndFor            
            \EndFor
            \EndFor
        \Return
	\end{algorithmic}
\end{algorithm}

The objective of a RL algorithm is to find the optimal policy to take actions, which maximizes the cumulative reward over time. The two fundamental methods in RL to find the optimal policy are the value based and policy based methods. The value based methods work by observing the 'Quality' or how good a particular state-action pair is, i.e., by using the Q function. In policy based methods, we find the optimal policy without calculating the Q function. Actor-critic methods take advantage of both the value and policy based methods in finding the optimal policy. In fact, they are proven to be able to overcome many shortcomings of vanilla policy gradient methods. Thus we form the basis of our scaling framework using the actor-critic algorithm. 

Actor-critic technique makes use of two neural networks, the actor network and the critic network. The actor helps find the optimal policy $\pi_{\theta}(a_{t}|s_{t})$, which leads to taking the best action in each state in order to achieve the desired objectives. The critic works in a feedback loop evaluating the policy generated by the actor, leading it to finding the best policy. In essence, the actor network is a policy network which uses a policy gradient method to find the optimal policy, while the critic network is a value network which is trained to estimate the state-value function, $v_{\pi}(s_{t}|\phi)$. $\theta$ and $\phi$ are the adjustable parameters of the actor and critic networks respectively.

Actor and critic networks learn by either maximizing their objective functions or by minimizing the loss functions. Accordingly, the actor learns the optimal policy by calculating the policy gradient, i.e., the gradient of the network and periodically updating the network parameter $\theta$ using gradient ascent (Eq.(\ref{eq:12})). 

\begin{equation}
\label{eq:12}
\begin{split}
    \nabla_{\theta}J(\theta) = \nabla_{\theta}log\pi_{\theta}(a_{t}|s_{t})A(s,a) \\
    &\  \hspace{-46mm} \theta = \theta + \alpha \nabla_{\theta}J(\theta)
\end{split}
\end{equation}

where $J(\theta)$ is the objective function which aims to increase the probability of occurrence of the actions which maximize the expected return of a given trajectory. As seen in Eq.(\ref{eq:12}) above, we calculate the policy gradient in the actor-critic methods using $A(s,a)$, the advantage function, hence the name Advantage Actor Critic (A2C). Expanding the advantage function;

\begin{equation}
\label{eq:13}
\begin{split}   
    \nabla_{\theta}J(\theta) = \nabla_{\theta}log\pi_{\theta}(a_{t}|s_{t})(r + \gamma V_{\phi}(s'_{t})- V_{\phi}(s_{t})) \\
&\ \hspace{-27mm} s.t.: (2), (3)
\end{split}
\end{equation}

Thus, the advantage function reveals how good action $a$ is compared to the average actions in state $s$. This essentially helps actor-critic methods to overcome inefficiencies of vanilla policy gradient algorithms by reducing the high variance of policy networks and stabilizing the model. Similarly, the critic learns by minimizing the loss of the critic network, i.e., the Temporary Difference (TD) error, which is the difference between the target value of the state $(r + \gamma V_{\phi}(s'_{t}))$ and the value of the state predicted by the network. During the course of training, the gradient of the critic network is calculated and the network parameter is updated using gradient descent (Eq.(\ref{eq:14})), thus allowing the critic to learn the actual state-value function.

\begin{equation}
\label{eq:14}
\begin{split} 
    J(\phi) = r + \gamma V_{\phi}(s'_{t})- V_{\phi}(s_{t}) \\
 &\  \hspace{-41mm} \phi = \phi - \alpha \nabla_{\phi}J(\phi)
\end{split}
\end{equation}

DRL techniques trained with a single agent have proven to be able to provide effective solutions for many single function scaling scenarios \cite{zhang2022adaptive}, \cite{somma2020less}, \cite{agarwal2021reinforcement}. But serverless platforms are usually multi-tenant environments with a number of deployed functions with various resource characteristics co-existing with each other. Further, these different functions have dynamically changing workload patterns, lowering the sample efficiency of many single agent RL solutions in the context of the multi-tenant scaling problem considered in our work. Thus in this work, we explore the applicability of the DRL technique A3C \cite{mnih2016asynchronous}, which employs several DRL agents who engage in learning in parallel, and aggregate the overall experience. The process of parallel learning helps explore the combination of state and action spaces much faster.

In A3C we work with two types of networks, the global network and the local or worker networks. Each worker agent interacts independently with its own copy of the environment, and shares the gathered experiences with the global agent asynchronously. Both the worker agents and the global agent follow an actor-critic architecture. Under A3C, in order to encourage sufficient exploration and reaching a global optimum, we add the term 'entropy' to the previously discussed (Eq.(\ref{eq:13})) actor loss, i.e.;

\begin{equation}
\label{eq:15}
\begin{split}   
    J(\theta) = log\pi_{\theta}(a_{t}|s_{t})(r + \gamma V_{\phi}(s'_{t})- V_{\phi}(s_{t})) + \beta H(\pi(s)) \\
\end{split}
\end{equation}

where $H(\pi)$ refers to the entropy of the policy while $\beta$ controls the significance of the entropy.
As discussed under section 4.1, we express our action space for scaling as a novel multi-dimensional discrete action space. We then adapt the technique described for discretized multi-dimensional action spaces in \cite{tang2020discretizing}, to design our actor network architecture.

We assume our normalized initial action space to be $A = [-1, 1]^{M}$, where $M$ represents the number of action dimensions. If we discretize each of these dimensions into K equally spaced actions, the set of atomic actions we get for each dimension $i$ is, $A_{i} = \{\frac{2j}{K-1}-1\}^{K-1}_{j=0}$. Then we present the distribution of action space as factorized across dimensions, in order to tackle the curse of dimensionality. As such, we consider a marginal distribution $\pi_{\theta_{i}}(a_{i}|s)$ for each dimension $i$, over the set of actions $a_{i} \in A_{i}$, where $\theta_{i}$ is the parameter of the distribution. Accordingly we get a joint discrete policy $\pi_{\theta}(a|s) = \prod^M_{i=1}\pi_{\theta_{i}}(a_{i}|s)$, where $\theta$ represents the parameter of the actor network which takes state $s$ as input. After layers of transformation, the network outputs the log probability $L_{ij}$ for the $j^{th}$ action in the $i^{th}$ dimension, where $i \in [1, M]$ and $j \in [1, K]$. Finally, for each dimension $i$, the K logits are combined with soft-max to derive the probability of choosing action $j$, i.e., $p_{ij} = softmax(L_{ij})$. Note that as per the scaling problem space defined in our work, $M = 3$ and K is chosen suitably for each action dimension. Each actor in our multi-agent framework follows this network architecture.

Algorithm \ref{algo:A3Calgorithm} presents the pseudo-code for the multi-agent scaling framework training process flow. We first initialize the global actor and critic network parameters which would be shared among and updated by the worker agents during the training process. Each worker would have its own copy of the environment and separate actor and critic networks (lines 3-6). At the start of each episode, workers reset their local environment. At each time step, the agent retrieves the state information with regard to the platform and the function to be scaled, and feed it to the local actor network. Next, the marginal probability distribution for each action dimension is used to determine the combined action for the current step (lines 11-12). Upon execution of the generated action, the environment transitions to a new state, and the agent receives a reward. All the transition information which includes the environmental state, executed action, awarded reward, and the next state are stored in memory (line 14). If the network update frequency or the maximum step count for an episode is reached, the agent starts the network parameter sharing and update process. First the advantage estimates, the loss and the network gradients are calculated for each transition stored in memory (lines 16-18). Then each worker agent asynchronously updates the global actor and critic network parameters using the calculated gradients. Finally, the local networks are updated with new weights pulled from the global model. After each network update, the local memory is cleared (lines 19-21). 

\section{Performance Evaluation}
\subsection{RL Environment Design and Implementation}

We implement a practical experimental serverless framework on the Melbourne Research Cloud \cite{Melbourn11:online} for preliminary data collection related to profiling serverless functions and also for deriving realistic system parameters exhibited during function executions. The testbed comprises of the Kubeless \cite{Kubeless28:online} open source serverless framework deployed on a 20 node Kubernetes \cite{Kubernet0:online} cluster on top of which the Prometheus \cite{Promethe59:online} monitoring tool is installed for monitoring cluster metrics. 
 
Following the architecture of this practical testbed, we have developed a simulation environment for serverless function execution in \textit{Python}, which also represents the system model presented in section III(A). This environment is integrated with Tensorflow-agents in the backend, which are developed using Keras and Tensorflow(TF) libraries. The key features of our developed simulator environment and the agent training process flow are summarized below:

\begin{enumerate}
\item Requests arriving at deployed function instances are loaded to an event queue in the order of arrival at the start of the simulation.
\item In the event that a suitable function instance is unavailable to accommodate an incoming request, the request is queued and subsequently dropped, after multiple scheduling retries at set time intervals.
\item Time steps for scaling decision making for each agent are scheduled at regular time intervals so that the agent's learned policy is capable of supporting proactive scaling of function resources independent of any workload specifics.
\item At each time step of the DRL agent, the serverless environment exposes the cluster state metrics which include the VM resource usage statistics and the workload nature of the function to be scaled.
\item After the execution of each of the combined horizontal and vertical scaling actions, the agent waits for a set time duration for the environment to reflect the action consequences, before deriving the step reward.
\item Each agent in the implemented multi-agent model, follows these steps in parallel, on copies of the same cluster environment.
\end{enumerate}

Although our implemented TF-agents are tasked with optimizing function performance and cluster resource cost, the developed simulation environment is capable of exposing monitoring metrics required for any other  extended objectives and facilitating training for continuous action spaces or modified DRL agent architectures. Our RL based serverless environment implementation with TF agents as the back end called 'Serverless\_DRL', is available as an opensource software \footnote{https://github.com/Cloudslab/Serverless\_DRL}.

\subsection{Experimental Settings}

\subsubsection{Cluster Setup}
Our simulated VM cluster comprises of 20 heterogeneous VMs of 4 different vCPU and memory configurations. The clock speeds of the CPU cores were set to be similar to that of VMs in AWS Lambda serverless platform as identified in \cite{wang2018peeking}. We use the AWS instance pricing of EC2 VMs (in Australia) \cite{AWSPrici23:online} closely matching the clock speed, vCPU and RAM configurations, as our pricing model. These cluster resource details are summarized in Table \ref{table:resourcedetails}. Our practical testbed too follows these VM cpu and memory configurations. We conduct our experiments under two scenarios, letting the multi-agent model to be comprised of 3 and 5 parallel actor-learners (agents) under each scenario, in order to observe the training time and data efficiency in state and action space exploration with more agents. Each individual agent works in a cluster environment of similar configuration as above during training.

\begingroup
\begin{small}
\begin{table}[!t]
	\caption{Worker Cluster Resource Details}
	\label{table:resourcedetails}
	\centering
    \resizebox{\linewidth}{!}{
        \begin{tabular}{l c c c c}
			\hline
			\\
			\multicolumn{1}{c}{\textbf{Instance Type}}
			&\multicolumn{1}{c}{\textbf{\centering vCPU cores}}
			&\multicolumn{1}{c}{\textbf{\centering Memory(GB)}}
			&\multicolumn{1}{c}{\textbf{Quantity}}
			&\multicolumn{1}{c}{\textbf{Price($\$$/hr)}}
			\\
			\hline
                m6g.medium & 1&4&5&0.048 \\
			t4g.large &2& 8& 5 &0.0848\\
			t4g.xlarge&4&16& 5 &0.1696 \\
			t4g.2xlarge &8&32&5&0.3392 \\
			\hline
	\end{tabular}}
\end{table}
\end{small}
\endgroup

\subsubsection{Workload Specifications}
\hfill

\indent\textbf{Serverless Applications}: We choose 12 benchmark applications from  ServiBench \cite{scheuner2022let} and FunctionBench \cite{kim2019functionbench} benchmark suites, which are formed of either a single or a chain of functions. Each of these applications have varying demands on CPU and memory resources based on their constituent functions. Thus their diverse sensitivities to different horizontal and vertical actions in the action space provide a good learning experience for the DRL agents. We use our practical setup for conducting function profiling for all the selected applications. An instance of each individual function is deployed on a VM in isolation and the JMeter \cite{ApacheJM57:online} load generation tool is used for sending a series of user requests to this instance. The results obtained from this tool and the cluster data recorded by Prometheus are averaged across multiple such workload executions to determine the resource consumption of a single function request ($Req^{k}_{cpu}(t) and Req^{k}_{mem}(t)$), standard response time ($r^{k}_{0}$) of a request and the instance creation time. $P^{k}_{cpu}$ and $P^{k}_{mem}$ for each function is initially set as the resources required to handle a defined number of requests. Table \ref{table:appdetails} captures the nature of these benchmark applications. 

\begingroup
\begin{tiny}
\begin{table}[!b]
	\caption{Serverless Application Details}
	\label{table:appdetails}
	\centering
    \resizebox{\linewidth}{!}{\begin{tabular}{l c c c}
			\hline
			\\
			\multicolumn{1}{c}{\textbf{Name}}
			&\multicolumn{2}{c}{\textbf{\centering Resource Sensitivity}}
			&\multicolumn{1}{c}{\textbf{\centering \# of Functions}}\\
			
			\cline{2-3}
			&CPU & Memory
			\\
			\hline
			Primary & High & High &1\\
			Float& High & High &1 \\
			Matrix Multiplication & High & High &1\\
			Linpack & High & High &1\\
			Load&low&low&1 \\
			Dd&High&Medium&1 \\
			Gzip-compression&High&Medium&1 \\
			Thumbnail Generator & Low&Medium&2\\
			Facial Recognition & Medium&Medium&5\\
			Todo API &Low&Low&5\\
			Image Processing & Medium&Medium&2 \\
			Video Processing &High&High&2\\
			\hline
	\end{tabular}}
\end{table}
\end{tiny}
\endgroup

\textbf{Workload Creation}: We leverage function traces from the publicly available data set from Microsoft Azure's Serverless Platform \cite{shahrad2020serverless} in order to derive request arrival patterns when creating the function workloads for both training and evaluating the DRL agents. In all the experiments, we maintain request arrival rates at 10-60 requests per second, maximum compute power and memory allocated to a single function instance at 1 vCPU core and 3GB respectively and the execution time of a request below 10 seconds. Accordingly, we analyse the Azure function data collected over a 24 hour period and filter a set of multi and single function applications with these characteristics and extract their request arrival patterns in the workload creation. For each function the per minute request arrival rates recorded in Azure traces for a given function is considered as a per second rate. In a given workload, the function arrival rates for a single function is fluctuated over time using these traces. Multiple such function request loads are combined to form a single workload. A workload consumed by a single agent is incorporated with traffic from no more than 4 functions at a time in order to maintain a sufficient load in the cluster for the training process. During each time step, the function subjected to scaling configuration changes is decided arbitrarily during the training process while the function with the highest RFRT is chosen during model evaluation, in order to attain optimum application performance. 

\subsubsection{Hyper-parameter Configurations} 
Hyper-parameters for the actor critic networks of each worker agent are decided on a trial and error basis. The discount factor is maintained at a lower value since the rapidly fluctuating nature of serverless workloads reduce the relevance of distant rewards towards current actions and thus a higher discount factor would force irrelevant information on the agent hindering the learning process. The learning rate for the actor and critic networks is maintained low enough so as not to cause a gradient blowup or lead to a sub-optimal solution too fast, and high enough so as the model converges with sufficient training. Each action dimension is discretized in to 11 actions as described under section IV(B). This was arrived at after  a series of initial experiments in order to maintain action space exploration costs at a manageable level while reaching good optimization levels for the target metrics. The maximum number of replicas for a single function at a time was restricted in order to suit the capacity of a 20 VM cluster while an upper limit of CPU scaling threshold was also set to ensure proactive scaling for all functions even at very low traffic levels.  The hyper-parameter settings for the A3C agents along with these environmental parameters in use for all the experiments are listed in Table \ref{table:hyperparameters}.  

\begingroup
\setlength{\tabcolsep}{2pt} 
\renewcommand{\arraystretch}{1} 
\begin{table}[!t]
\caption{Hyper-parameters Used for DRL Model Training}
\label{table:hyperparameters}
\small
\begin{tabular*}{\columnwidth}{p{0.6\columnwidth}|p{0.4\columnwidth}}
\hline
\textbf{Parameter} & \textbf{Value} \\
\hline
\textbf{General} \\
Optimization parameter ($\beta$) & [0.0, 0.25, 0.50, 0.75, 1.00] \\
Maximum number of concurrent replicas of a function & 80\\
Maximum pod CPU utilization for horizontal scaling & 90\% \\
\hline
\textbf{Neural network parameters}\\
Discount factor ($\gamma$) & 0.6 \\
Learning rate ($\alpha$) & 0.0001 \\
No. of input layers & 1 \\
No. of output layers & 1 \\
No. of hidden layers & 2 \\
No. of neurons in each hidden layer  & 150 \\
Optimizer & Adam \\
Network update frequency & 30\\
Action space size for each dimension & 11 \\
\hline
\end{tabular*}
\label{tab2}
\end{table}
\endgroup

\subsection{Performance Metrics}
We use three metrics to evaluate the effectiveness of our solution noted below:

\noindent
\begin{enumerate}
\item\textbf{Average Relative Application Response Time ratio (RART)}: The sum of the average, relative response times of all the applications in a workload during an episode, divided by the number of applications, calculated using Eq.(\ref{eq:7}). 

\noindent
\item\textbf{Request Failure Rate (RFT)}: The ratio of the number of dropped function requests to the total number of requests received in an episode.

\noindent
\item\textbf{VM Usage Cost}: The cost of maintaining the VMs active during an episode. The calculation of this metric is as in Eq.(\ref{eq:9}). 
\end{enumerate}

\subsection{Baseline Scaling Techniques}
We use four baseline scaling techniques to compare the performance of our proposed solution.

\noindent
\textbf{DQN}: We use the value based DRL algorithm Deep Q Learning to arrive at a solution for the function scaling problem. Here we consider each combination of the actions from the three discretized action spaces for horizontal scaling, CPU and memory vertical scaling as a compound action that the agent chooses. Due to the state action space explosion that results from combining actions in this way, we limit the granularity of action discretization to four actions per dimension resulting in 64 compound actions in total.

\noindent
\textbf{Knative}: The opensource serverless platform Knative \cite{Aboutaut63:online} allows users to set a target pod concurrency value, limiting the number of concurrent requests handled by a function instance. Further a target utilization value is set determining the actual percentage of the target that we should meet. Horizontal scaling of functions is triggered in order to maintain the set level of request concurrency and the target utilization (we set these at 4 and 75\%).

\noindent
\textbf{Kube-cpu}: Kubernetes default horizontal pod auto-scaler scales function instances based on a set threshold on function level resource usage metrics. Here we consider scaling function replicas in order to maintain the average CPU utilization across all instances of a function at or below the set value (we set the CPU utilization threshold at 50\%). 

\noindent
\textbf{OpenFaaS}: The opensource serverless platform OpenFaas \cite{Autoscal28:online} offers a mix of three modes of scaling to be used based on the function requirements. For long-running functions which can handle only a limited number of requests at a time, the 'capacity' mode triggers scaling based on the number of in-flight requests (we consider a threshold set at 4). For functions which execute quickly and have a high throughput, the 'rps' mode enables scaling based on the number of requests per second completed by a function replica (threshold set at 8). All the other workloads which do not support the 'capacity' and 'rps' scaling profiles use the 'cpu' mode which triggers scaling based on the average CPU utilization (set at 50\%) across pods, similar to Kube-cpu. We dynamically decide on the scaling mode used for each function type based on their execution times and request rates. Accordingly, functions with execution times greater than 2 seconds are scaled using the 'capacity' mode, functions with less than 2 seconds execution time and request rate higher than 20 requests per second at the time, use 'rps' mode and the rest are scaled using the 'cpu' mode. 

\begin{figure*}[!t]
    \centering
    \includegraphics[width=0.55\textwidth, height=0.4cm]{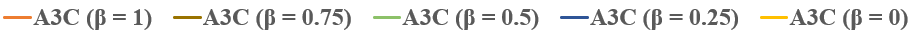}\\
    \subfloat[Reward Convergence]{\includegraphics[width=.245\textwidth,
    height=3.5cm]{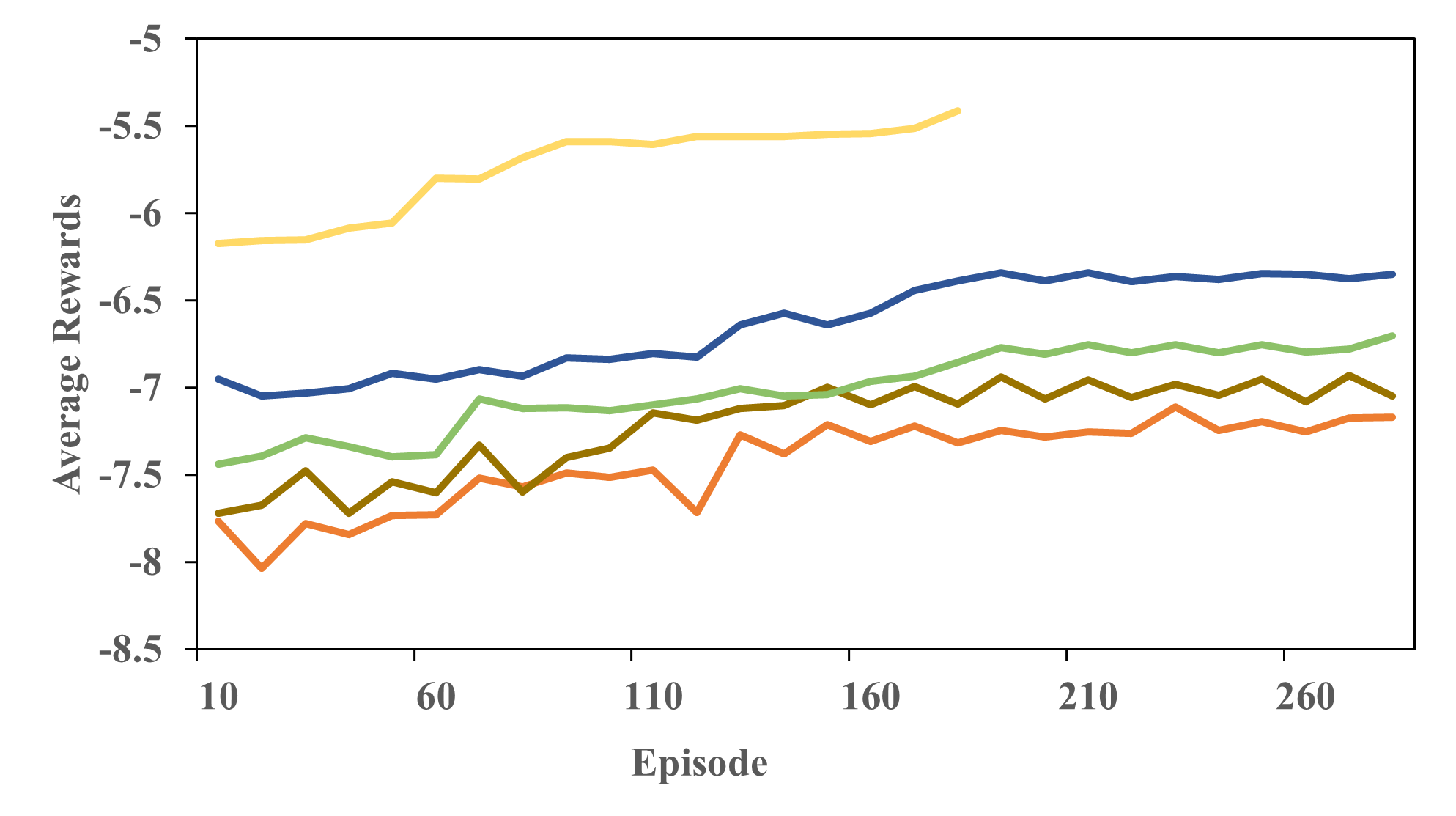}}
    \subfloat[Average Relative Function Response Time (RFRT)]{\includegraphics[width=.245\textwidth, height=3.5cm]{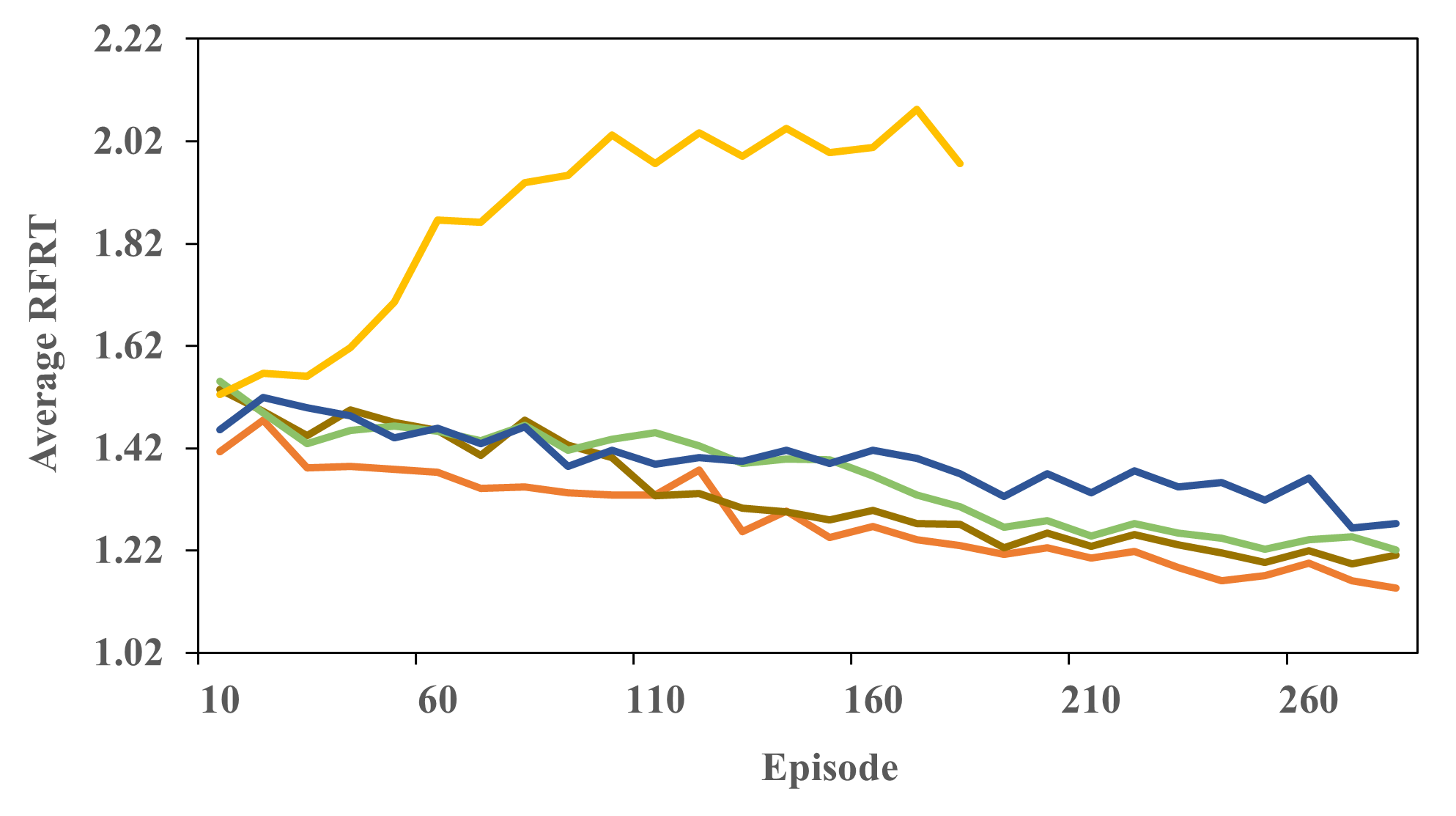}} 
    \subfloat[Request Failure Rate (RFR)]{\includegraphics[width=.245\textwidth, height=3.5cm]{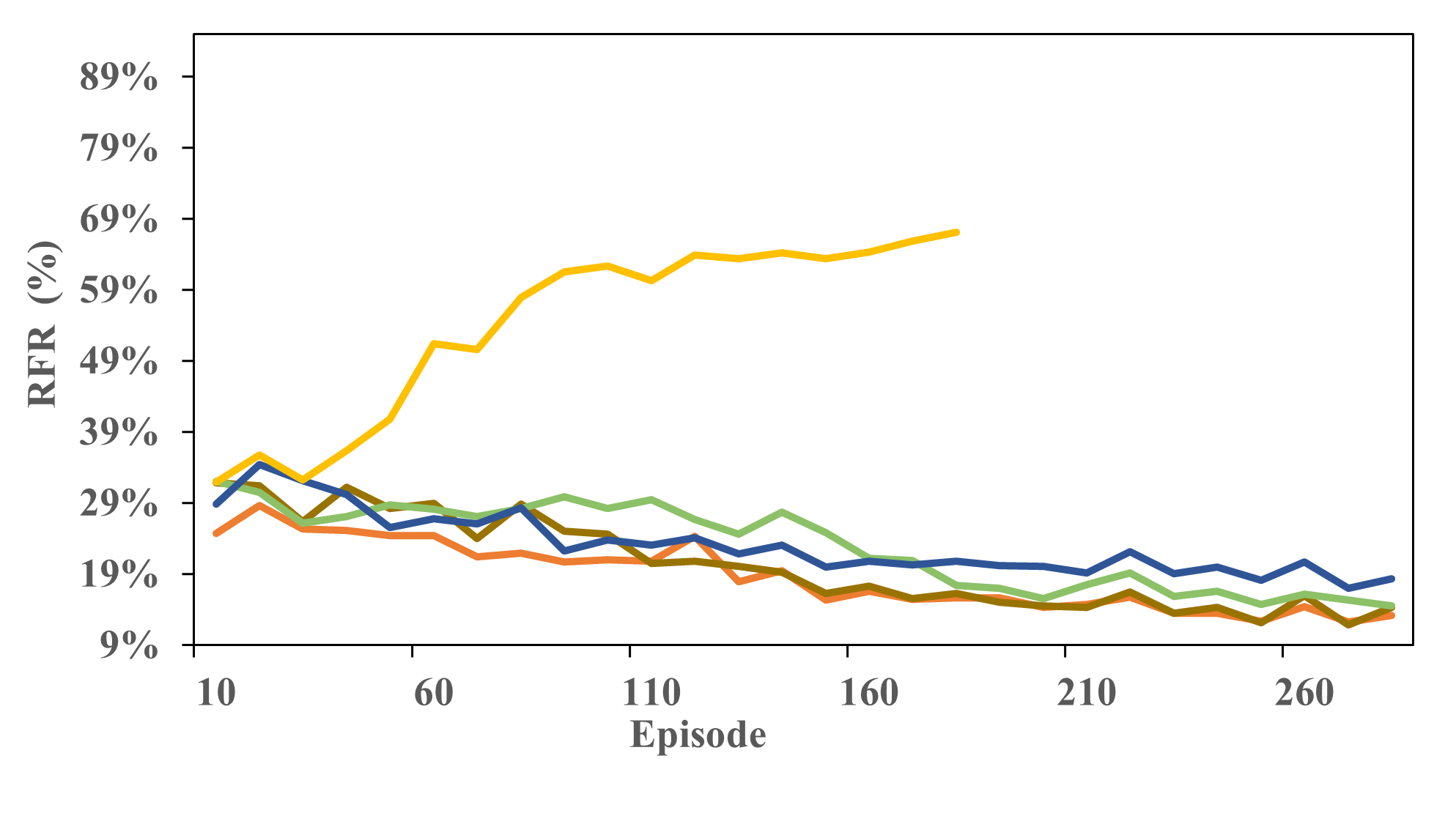}}
    \subfloat[Provider VM Cost]{\includegraphics[width=.245\textwidth, height=3.5cm]{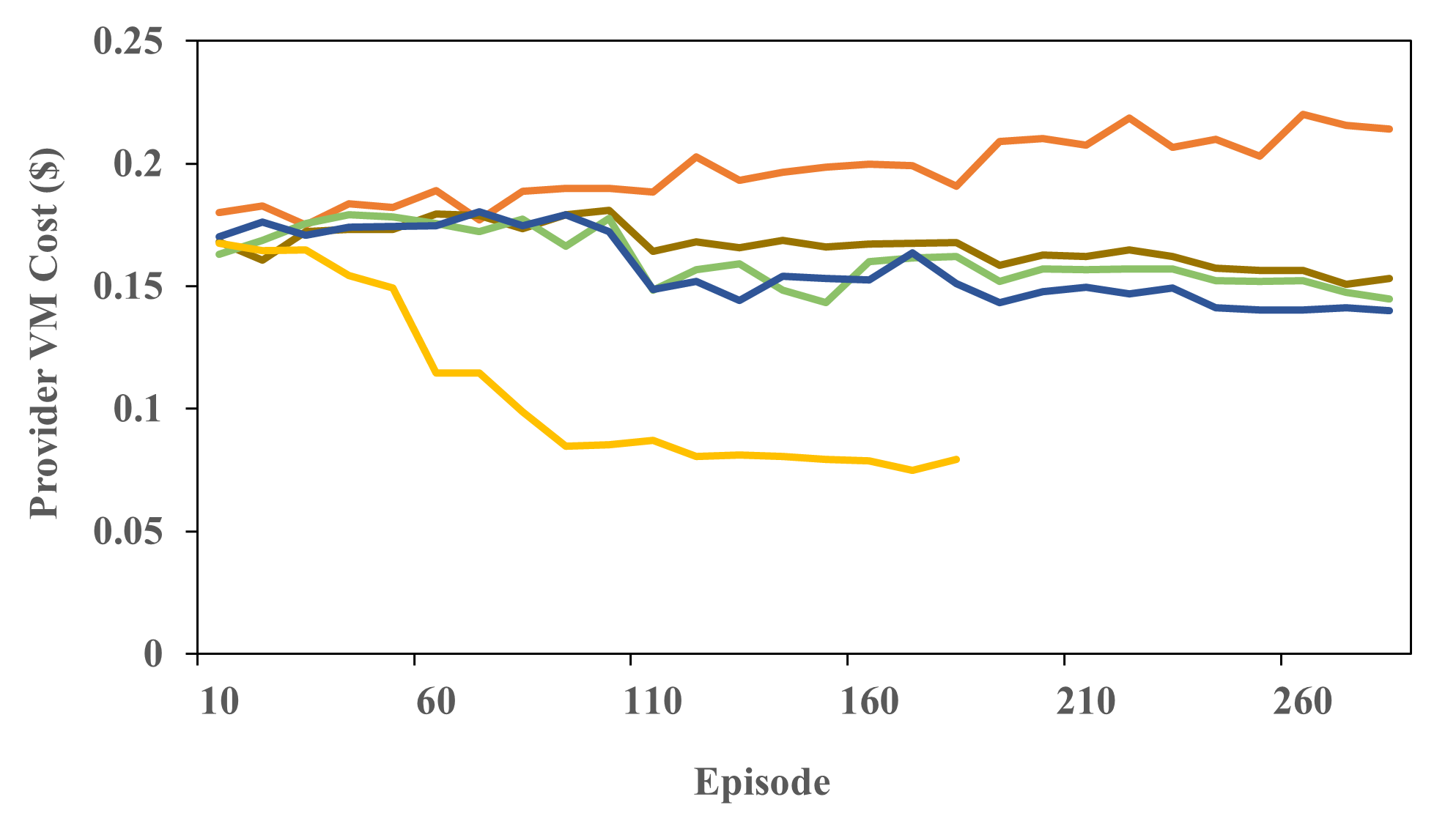}}
    \caption{Training progress of the 3 worker A3C models in terms of reward, average RFRT, request failure rate, and the total VM cost}
    \label{fig:convergence-3}
\end{figure*}

\begin{figure*}[!t]
    \centering
    \includegraphics[width=0.55\textwidth, height=0.4cm]{figures/legend1.PNG}\\
    \subfloat[Reward Convergence]{\includegraphics[width=.245\textwidth,
    height=3.5cm]{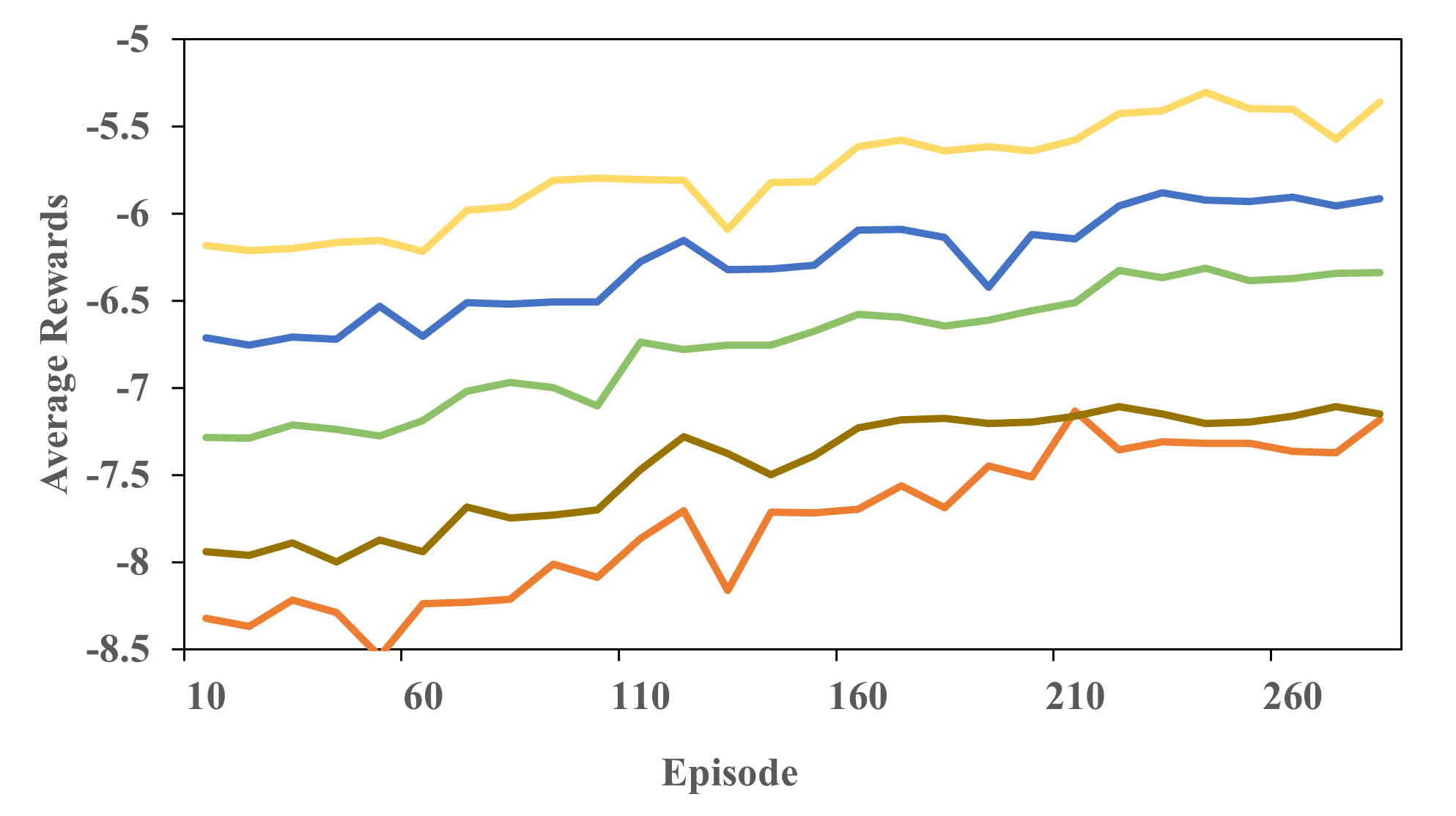}}
    \subfloat[Average Relative Function Response Time (RFRT)]{\includegraphics[width=.245\textwidth, height=3.5cm]{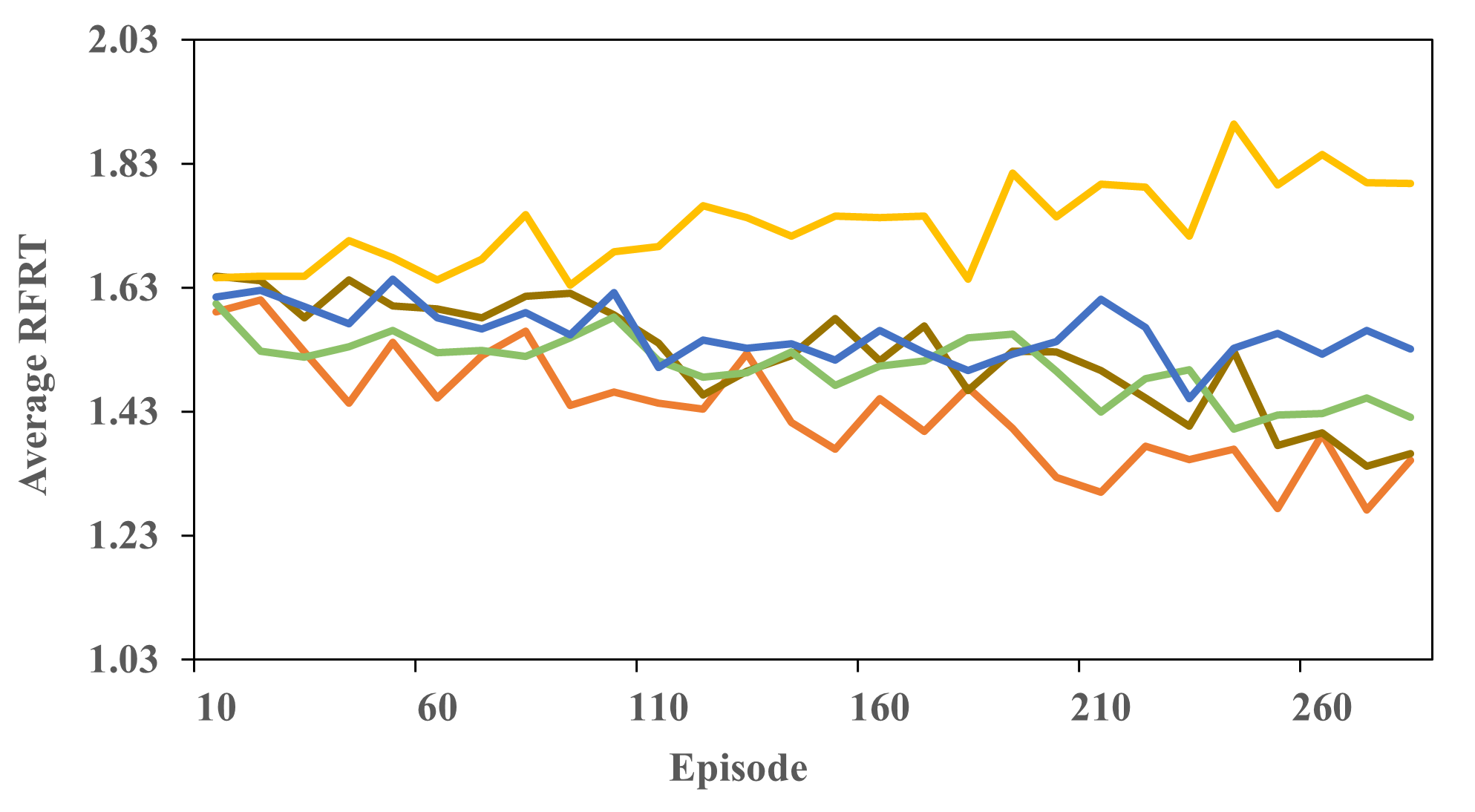}} 
    \subfloat[Request Failure Rate (RFR)]{\includegraphics[width=.245\textwidth, height=3.5cm]{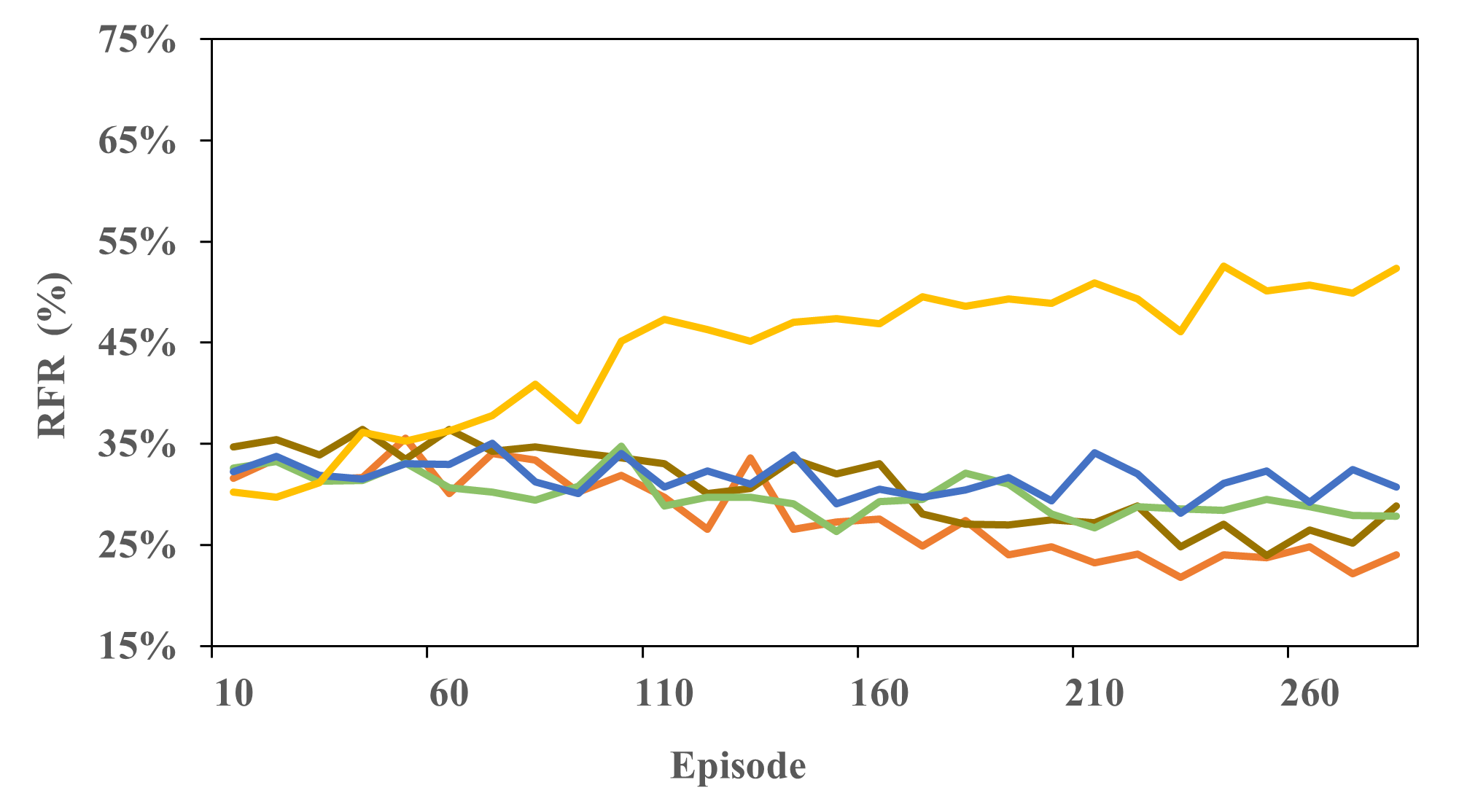}}
    \subfloat[Provider VM Cost]{\includegraphics[width=.245\textwidth, height=3.5cm]{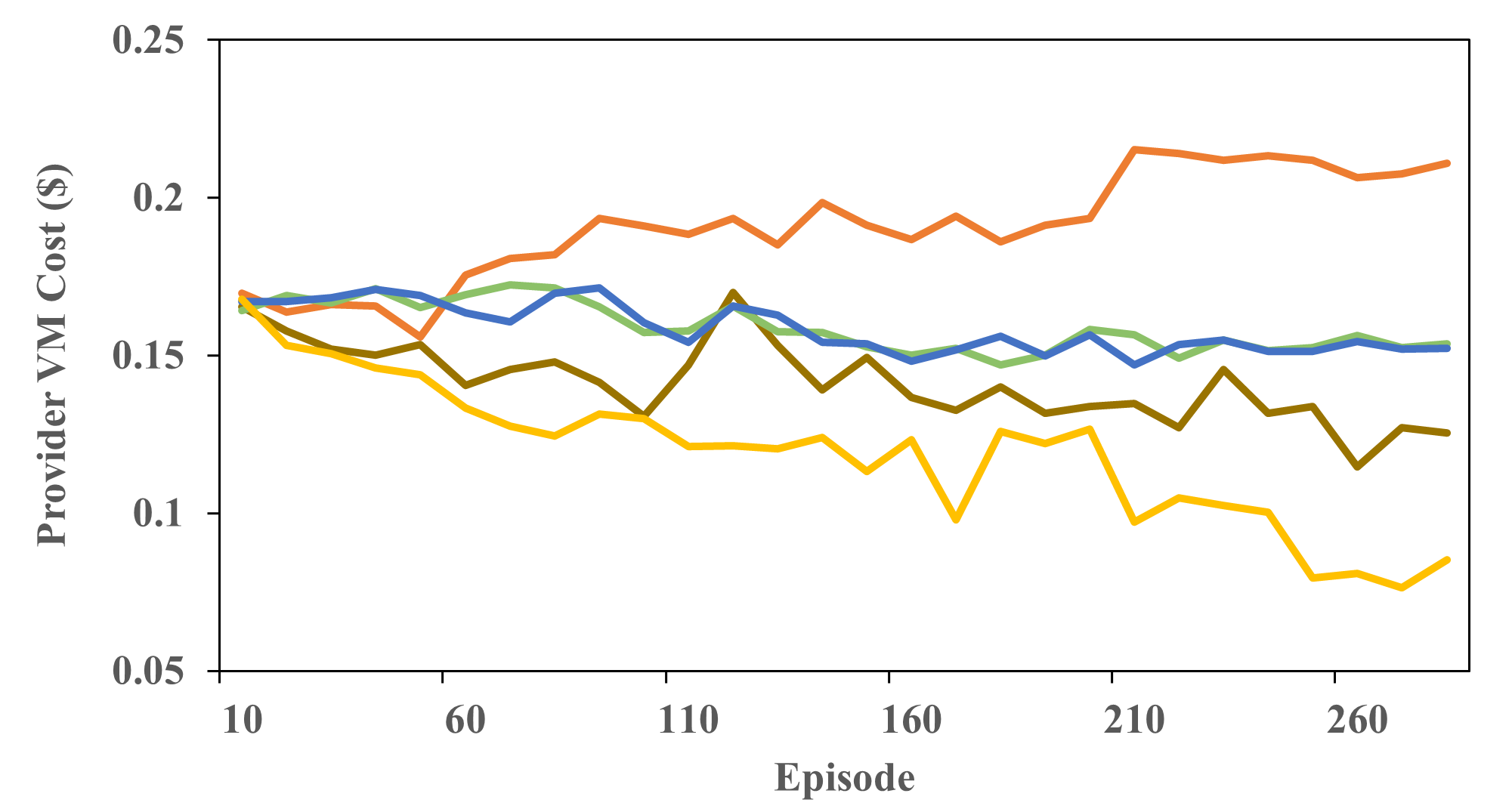}}
    \caption{Training progress of the 5 worker A3C models in terms of reward, average RFRT, request failure rate, and the total VM cost}
    \label{fig:convergence-5}
\end{figure*}

\subsection{Convergence of the DRL Model}

Under each of the multi-agent model scenarios comprising of different number of parallel workers, we train 5 variations of the A3C model considering the significance given to each optimization objective. We use the $\beta$ parameter to signify the priority assigned to each objective in the agent reward structure. Accordingly, $\beta = 1$ refers to a $100\%$ focus on improving function performance while $\beta = 0$ indicates model training leading to infrastructure cost optimization only. The graphs in Figs. \ref{fig:convergence-3}  and \ref{fig:convergence-5} display the training progress achieved over time as the agents gradually learn to optimize the cumulative reward over an episode. The progress is demonstrated in terms of the episodic reward and the target optimization metrics themselves, i.e: average relative function response time (RFRT), request failure rate (RFR) and the VM cost over each episode. Note that the marked values on the graphs are averaged values over 10 episodes for clarity in presentation.

In the first scenario with 3 actor-learners working in parallel, we train the model with 9 different functions, deployed in the cluster in total. These are chosen to be a mix of functions that are common to multiple applications from Table \ref{table:appdetails}. In the next scenario, we employee 5 actor-learners in the learning process, in order to analyse the achieved efficiency in state space exploration with more worker agents. In this second set of experiments, we deploy 15 different functions altogether in the cluster and observe the agent behavior leading to model convergence.

Figs.  \ref{fig:convergence-3}(a) and \ref{fig:convergence-5}(a) show the convergence of the episodic reward under each scenario with varying $\beta$ parameters. As seen from the graphs, all of the models achieve convergence around the $300^{th}$ iteration, despite the considerably expanded state space size in the second scenario. This is due to the speed-up in data exploration achieved by having more workers learning in parallel, which results in the global model reaching its maximum optimization levels faster, effectively improving both time and data efficiency. Fig. \ref{fig:convergence-3}(a) also shows that when $\beta=0$, the model convergence happens relatively faster compared to other scenarios in the set of 3 worker models. Since  $\beta=0$ only incentivizes reducing the VM cost, the agent seems to easily learn to take actions that lead to maintaining the lowest number of replicas possible in the cluster while also limiting their CPU and memory capacities. On the other hand, improving function performance is not as straightforward for the agent to learn, since expanding the pool of function replicas or vertically scaling pod resources would not always lead to the optimum solution. This is because while horizontal scaling creates new resources for request execution, it also adds a resource set up delay which causes increased latency and request failures. Thus a more intelligent strategy needs to be learned depending on the cluster state at each scaling step.

Figs. \ref{fig:convergence-3}(b), \ref{fig:convergence-3}(c), \ref{fig:convergence-3}(d) and \ref{fig:convergence-5}(b), \ref{fig:convergence-5}(c), \ref{fig:convergence-5}(d) represent the average function latency, failure rates and the VM costs incurred over the corresponding training episodes. The  $\beta=1$ graphs in both \ref{fig:convergence-3}(b) and \ref{fig:convergence-5}(b) maintain a steady decrease in RFRT with each iteration. The $\beta=0.75$, $\beta=0.5$, $\beta=0.25$ graphs too show a gradual decrease in function response time latency since they too are partially motivated to improve function performance in the reward. But understandably, they converge at a higher latency level than when solely focused on optimizing this feature alone. The $\beta=0$ models on the other hand display a completely opposite trend of increasing latency with each iteration as they simply target resource efficiency only, and this easily compromises the performance parameter. The RFR graphs too display a similar trend in all the scenarios. The VM cost graphs show a clear decrease in VM cost over time when $\beta=0$. The $\beta=0.25$, $\beta=0.5$, and $\beta=0.75$ graphs too show a moderate decline in overall cost for the provider with time. The $\beta=1$ model converges at a high VM cost as expected, but here we observe considerably lesser prominence than the decline in function performance we earlier saw with the $\beta=0$ model. A probable cause for this behavior is likely to be the indirect effect that actions leading to improved performance seem to have on enhancing resource efficiency too.

\subsection{Analysis of Model Performance on the Evaluation Data Sets}

\begin{figure*}[!t]
    \centering
    \includegraphics[width=0.95\textwidth, height=0.4cm]{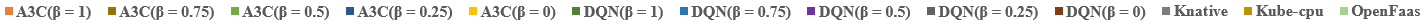}
    \subfloat[Average Relative Application Response Time (RART)]{\includegraphics[width=.49\textwidth, height=4cm]{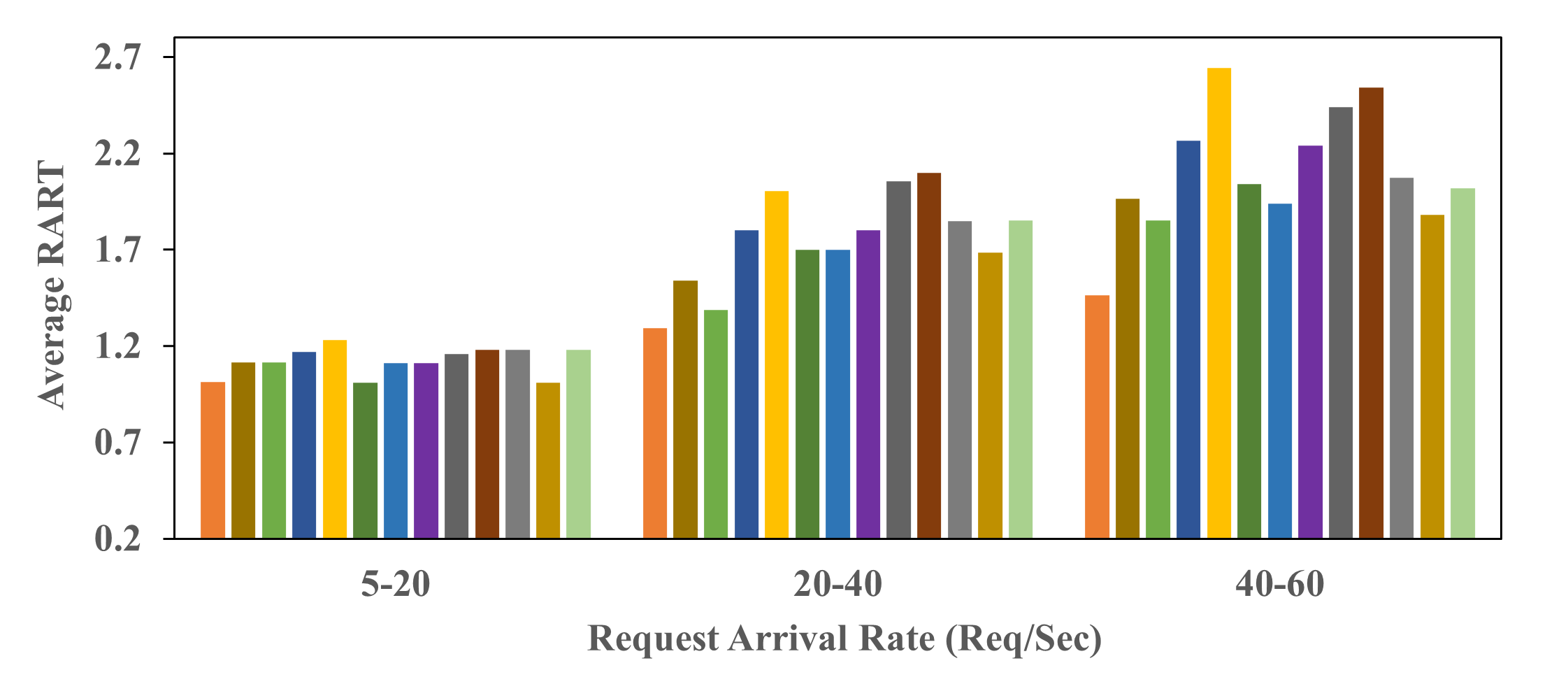}}\hfill
    \subfloat[Request Failure Rate (RFR)]{\includegraphics[width=.49\textwidth, height=4cm]{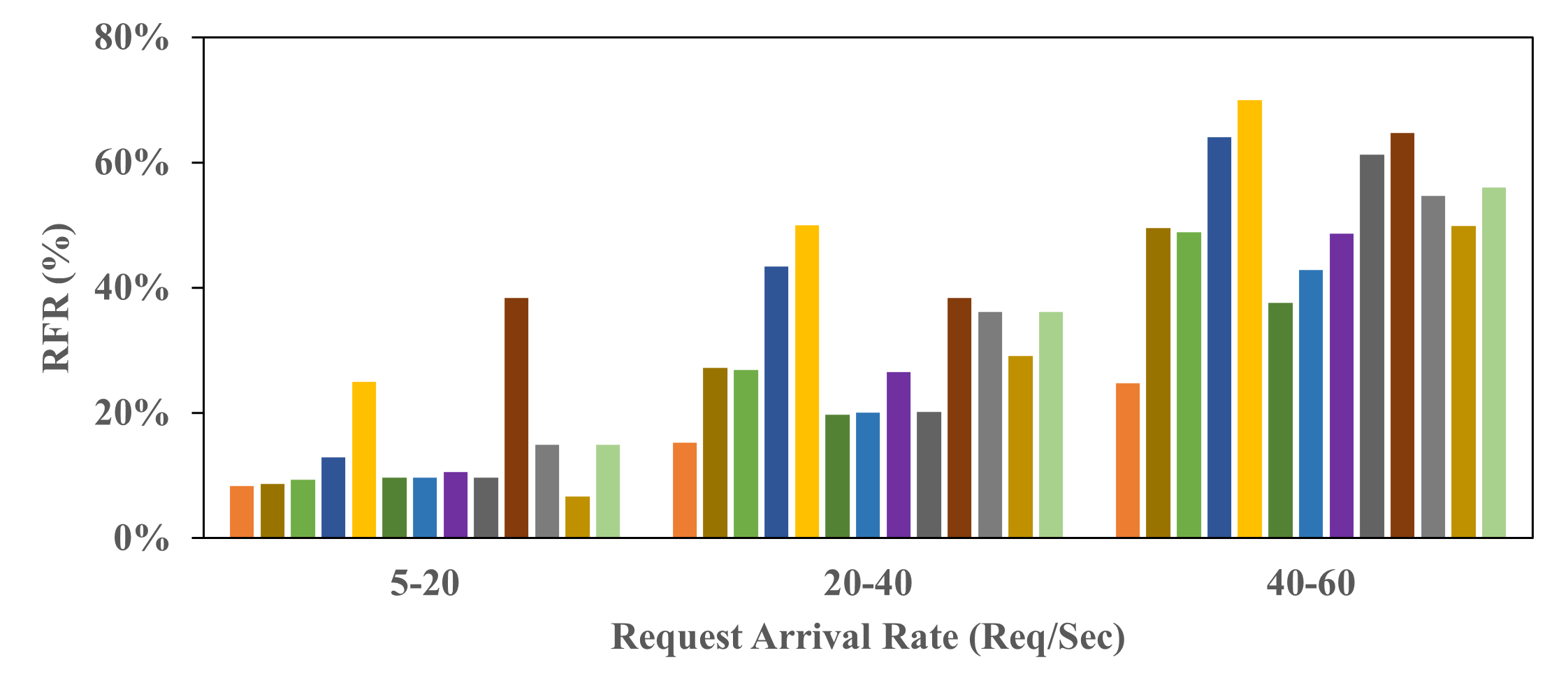}} \\
    \subfloat[Provider VM Cost]{\includegraphics[width=.49\textwidth, height=4cm]{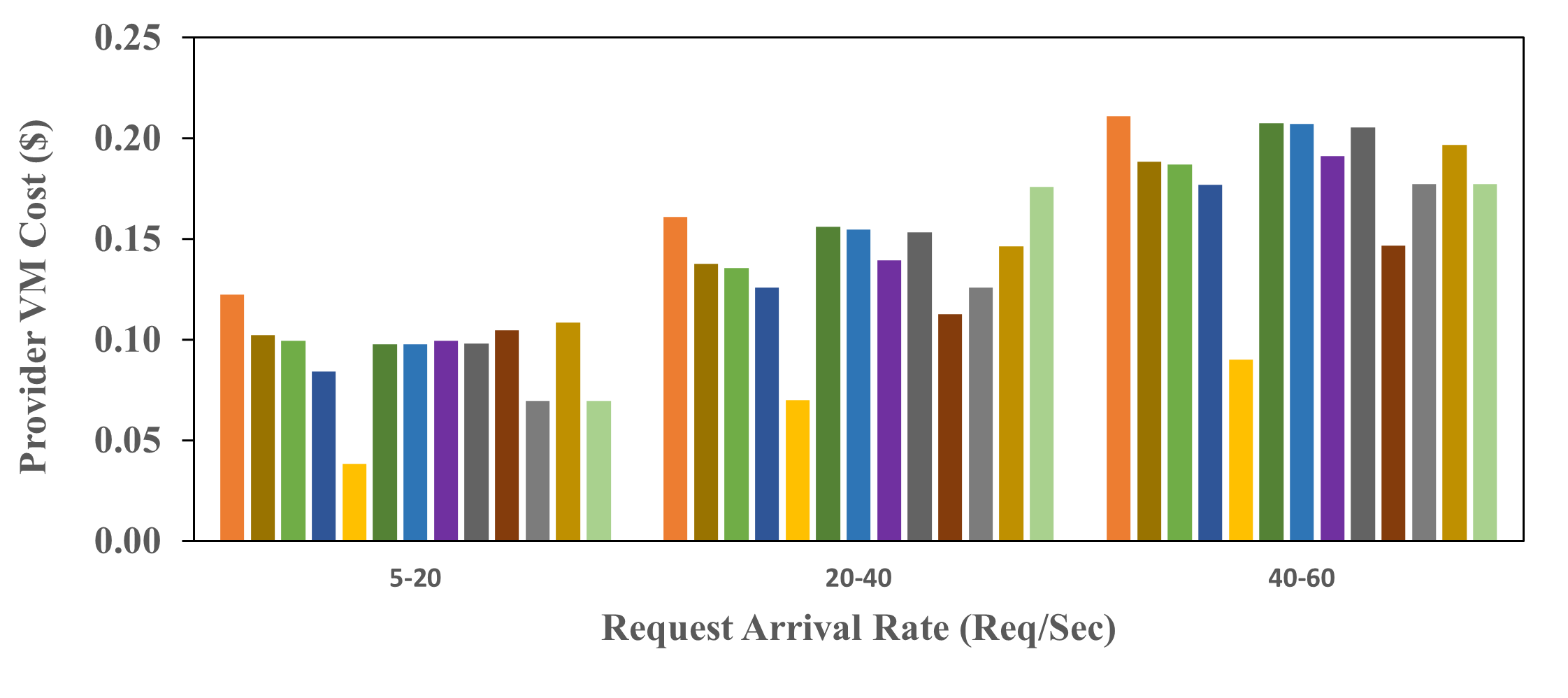}}\hfill
    \caption{Comparison of the Average RART, RFR and provider VM cost in the system during an episode, by the 3 worker A3C models and the baseline algorithms}
    \label{fig:evaluation3}
\end{figure*}

The performance of our trained multi-agent models is evaluated and discussed mainly in terms of our target optimization objectives of serverless application performance and resource cost efficiency. We extract 1800 function traces in total from Azure function traces using the procedure described in section V(B)(2) in creating the evaluation data set. Our model evaluation is conducted under three request traffic levels as 5-20, 20-40 and 40-60 requests per second, for both the 3 and 5 parallel agent scenarios. Accordingly, for both these scenarios, we create 60 workloads each for the 3 load levels, i.e. a total of 360 workloads. When creating each workload for evaluating the variations of the models trained with 3 agents, we include simultaneous user requests from 5 different serverless applications (single and multi-function) created using the  9 functions used during the training process. Similarly workloads are created for the 5 agent models incorporating requests from applications created using 15 different functions. Further, the request arrival rates for a single application are varied over time in a given workload, each of which spans over five minutes.

Fig. \ref{fig:evaluation3} and \ref{fig:evaluation5} demonstrate the performance of our A3C models against the baseline scaling techniques under the two agent scenarios. Each bar graph corresponds to the achieved performance metric derived by averaging over the 60 workload runs under each load level.

\subsubsection{Evaluation of application performance}
\begin{figure*}[!t]
    \centering
    \includegraphics[width=0.95\textwidth, height=0.4cm]{figures/legend2.PNG}
    \subfloat[Average Relative Application Response Time (RART)]{\includegraphics[width=.49\textwidth, height=4cm]{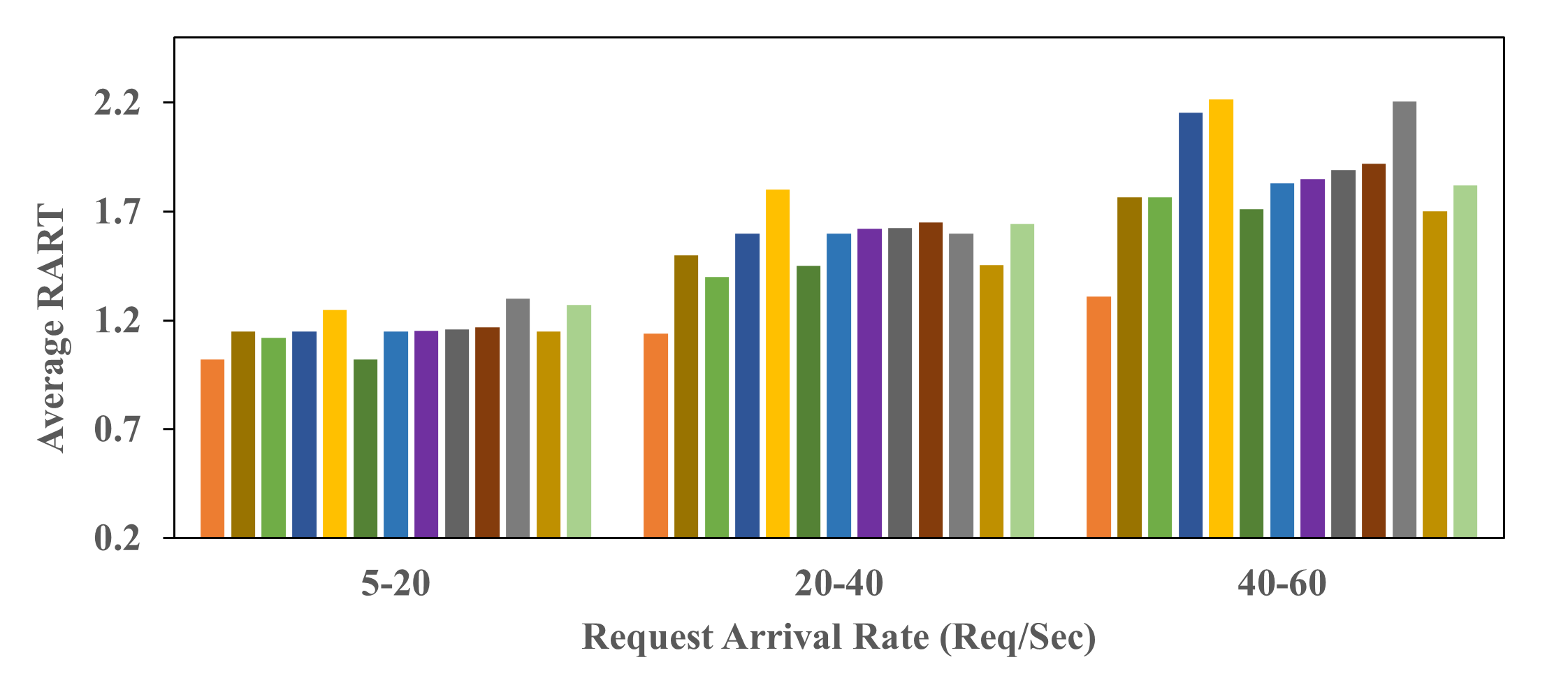}}\hfill
    \subfloat[Request Failure Rate (RFR)]{\includegraphics[width=.49\textwidth, height=4cm]{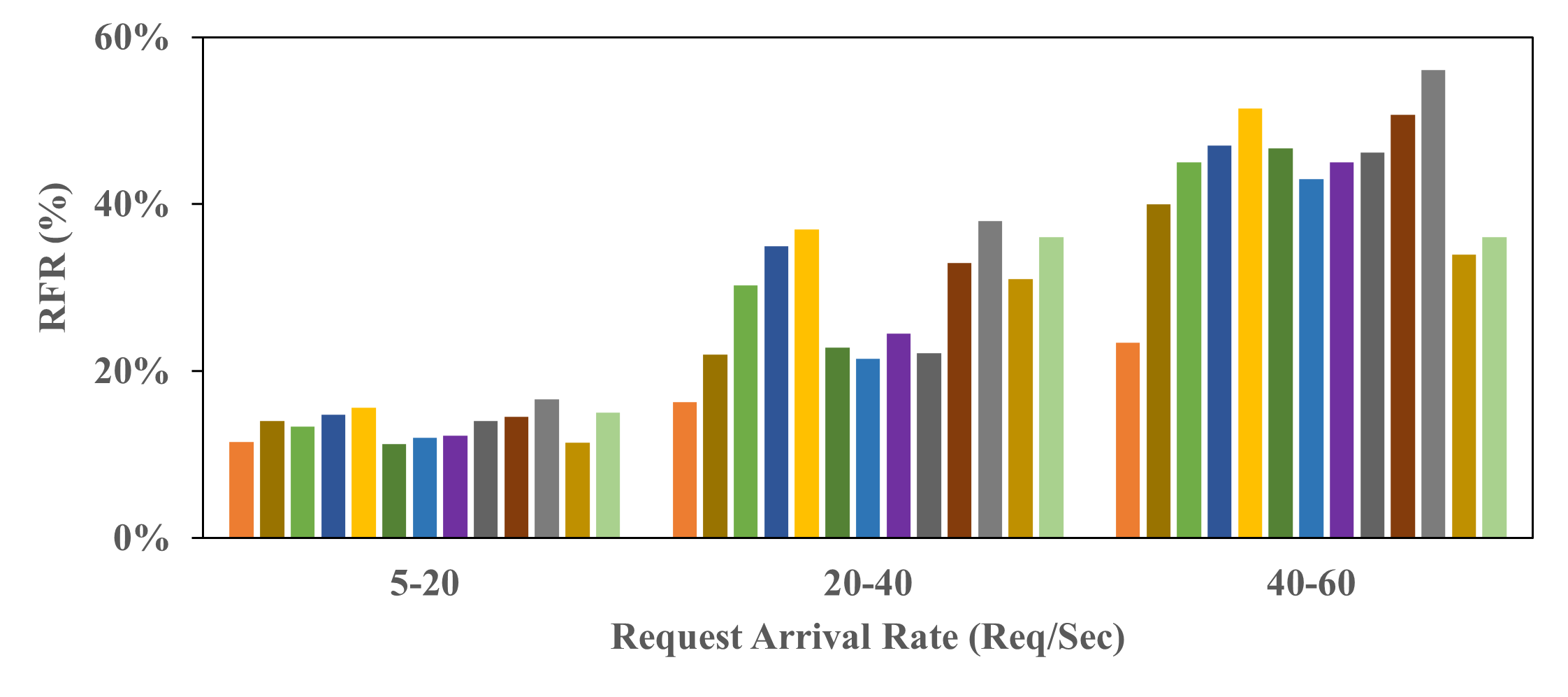}} \\
    \subfloat[Provider VM Cost]{\includegraphics[width=.49\textwidth, height=4cm]{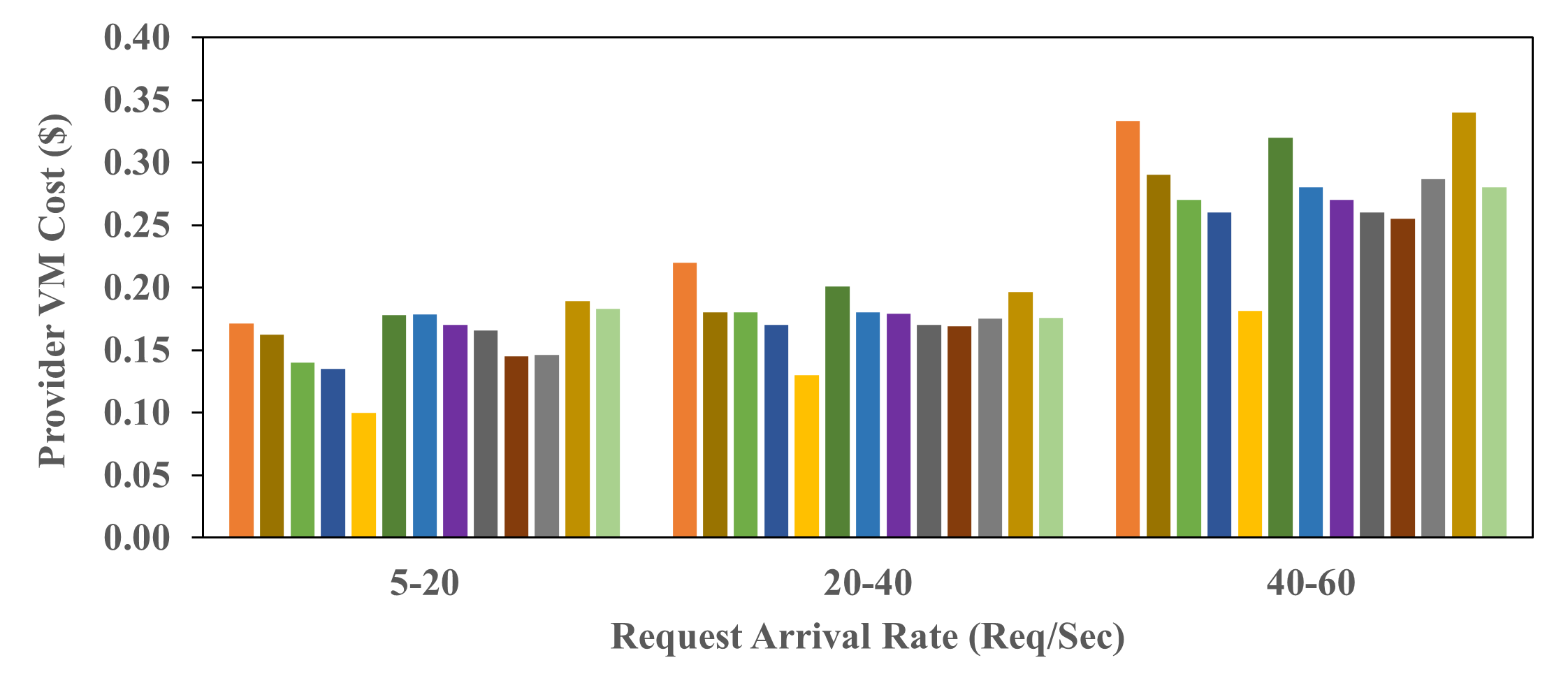}}\hfill
    \caption{Comparison of the Average RART, RFR and provider VM cost in the system during an episode, by the 5 worker A3C models and the baseline algorithms}
    \label{fig:evaluation5}
\end{figure*}

Application performance is evaluated in terms of RFRT and RFR performance as shown in graphs \ref{fig:evaluation3}(a), \ref{fig:evaluation5}(a) and \ref{fig:evaluation3}(b), \ref{fig:evaluation5}(b). Overall we can see that the latency and request failure rates increase gradually with the rise in request rates due to increased wait times for request executions arising from resource limitations in the cluster. Also, it is evident from the plotted graphs that the behavior of the models trained using both 3 and 5 actor-learner architectures, is similar in most aspects and thus our discussion below would entail a common analysis for both scenarios for the most part.

At the lowest traffic level of 5-20 req/sec, we do not observe a significant improvement from the A3C($\beta=1$) model compared to the rest, where the DQN($\beta=1$) and Kube-cpu models exhibit almost similar or better performance. This is because, at lower traffic levels, the cluster is less congested and thus an intelligent function scaling strategy adds less value to overall performance. However, at high $\beta$ values, the A3C as well as the DQN models show better function performance as their learned policy favors performance more than cost. 

As request rates increase to 20-40 req/sec, a more distinct performance upgrade is seen to be achieved by the trained models. In both the graphs for RFRT, \ref{fig:evaluation3}(a) and \ref{fig:evaluation5}(a) and for RFR, \ref{fig:evaluation3}(b), \ref{fig:evaluation5}(b), we observe the best performance from the A3C($\beta=1$) model. The A3C model outperforms the next best performing baseline by up to 23\% in RFRT and 24\% in RFR. Since our models in this case are purely rewarded for better function performance during the training process, they learn to maintain lower cpu thresholds for function scaling, leading to more proactive instance creation. Further, when an existing instance is reaching its maximum utilization levels, the agent learns to vertically scale its capacity after which it could immediately accommodate more requests without any additional wait times. In this process, the agent also learns to weigh between horizontal and vertical scaling as although vertical scaling expands capacity immediately, it limits future resource expansions. Thus if the current cluster load could sustain some delays in resource creation without excessive request failures, horizontal scaling could lead to long term performance benefits. The DQN($\beta=1$) model exhibits next best performance as it too follows an intelligent scaling policy in contrast to other baselines. However, the DQN model lacks the fine grained learning capability of the A3C model for many reasons. As a single agent model, it lacks the state space exploration capability even when trained for longer periods of time as we observed during model training. Also, since we had to create compound actions out of the 3 action dimensions with high level discretization, the extensiveness of action space exploration too was far weaker compared to the A3C model. The Knative, Kube-cpu and OpenFaas techniques which follow fixed threshold base scaling, do not perform well in a congested resource constrained cluster. They apply a blanket threshold for all the application functions facing varying request rates, which lead to increasingly poor performance as the cluster load rises. 

At 40-60 req/sec we see even more distinguished performance improvements in the A3C models with high $\beta$ values, with up to 34\% reduction in request failures when $\beta=1$. We also note that at times, the A3C($\beta=0.5$) model shows slightly better latency performance than the A3C($\beta=0.75$) model under high traffic levels. When the agent is rewarded equally to  improve both latency and cost ($\beta=0.5$), it has indirectly resulted in better function latency than when focused more on latency itself. This is because, scaling actions which lead to efficient cluster resource usage could also result in reduced request wait times and thus latency, which is an added advantage. The DQN model too shows similar behavior in the latency and request failure graphs for DQN($\beta=1$) and DQN($\beta=0.75$).  The  A3C($\beta=0$) models show worst performance in terms of application performance.

\subsubsection{Evaluation of resource cost efficiency}


Resource cost efficiency is evaluated in terms of the cost incurred by the provider to maintain the VMs while they contain running function instances. The overall cost of infrastructure increases as the load levels rise.

In contrast to our observations for latency performance at lower traffic levels, we see clear cost improvements of upto 45\% in the  A3C($\beta=0$) models trained for that purpose. This is because with lesser load, if cost is not a concern (i.e. at higher $\beta$ values), horizontal scaling is encouraged and the cluster could maintain a lot of idling instances. This leads to high VM maintenance costs. On other hand, where resource efficiency is rewarded, the agent learns to take vertical scaling actions more, which leads to higher utilization levels for the active VMs. Subsequently, any idling VMs could be switched off, which saves resource costs. Although not as efficient, the DQN models too show a decreasing trend in cost with $\beta$ at low load levels, in the second scenario (Fig. \ref{fig:evaluation5}(c)), which has higher multi-tenancy in the cluster with more applications. Kube-cpu scaling style triggers proactive horizontal scaling without a deeper understanding on the workload patterns, thus leading to large resource inefficiencies.

At 20-40 and 40-60 load levels too we observe a significant improvement in our A3C($\beta=0$) model, although the opportunity for gaining a huge resource efficiency level reduces as the cluster utilization levels increase. As expected , the next best performance is seen in the DQN($\beta=0$) model, as it closely follows the reward structure of the A3C model, falling only short of the state and action space exploration capabilities of the actor-critic architecture. A3C($\beta=1$), DQN($\beta=1$) agents exhibit worst performance in terms of cost, closely followed by the Knative, Kube-cpu and OpenFaas techniques which are unable to handle complex load scenarios to achieve a particular target.

\section{Conclusions and Future Work}
The abstract form of application resource management in serverless computing completely relieves the end users from operational responsibilities. However, cloud providers are still in the process of developing the best strategies to fulfill this new set of responsibilities. As such, attaining an optimum level of scaling for function resources of different applications is still a challenge requiring attention.

In this paper, we proposed a DRL based adaptive solution using the actor-critic architecture, for taking the horizontal and vertical resource scaling decisions for applications in a multi-tenant serverless environment. A successfully scaled application satisfies both the application owner and the infrastructure provider. Accordingly, application performance and the infrastructure maintenance cost for the provider, were considered as our target optimization objectives for DRL model training. Our solution offers flexibility for prioritizing either of these objectives, depending on the user requirements. We conducted and presented details of two sets of experiments in order to observe data and time efficiency improvements achieved, when using different numbers of parallel actor-learners in training our A3C model. Our trained DRL agents were able to take effective scaling decisions for functions deployed in serverless platforms, which led to reduced application latency, request failures and provider side resource wastage. We employed a trained DQN model, along with other baselines to evaluate our presented solution. The results obtained show that our presented intelligent scaling solution vastly benefits all user categories in meeting their objectives.

As part of future work, we plan to explore the efficiency of integrating a time-series regression model with the developed DRL agent architecture for function scaling, in order to forecast request traffic patterns. Further we aim to develop a framework for performance and cost modeling of functions, incorporating the performance data obtained in our scaling experiments by varying the allocated resources to function replicas.

\bibliographystyle{IEEEtran}
\bibliography{serverless.bib}

\begin{IEEEbiography}[{\includegraphics[width=1in,height=1.25in,clip,keepaspectratio]{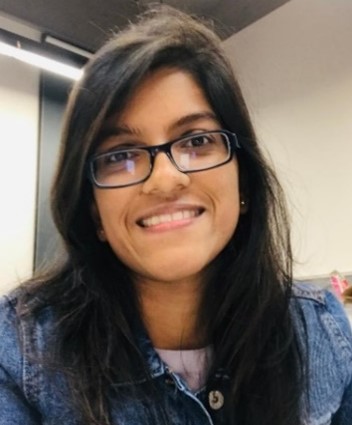}}]{Anupama Mampage}
 is a PhD student at the Cloud Computing and Distributed Systems (CLOUDS) Laboratory, Department of Computing and Information Systems, The University of Melbourne, Australia. She received her BSc Engineering (Hons) degree, specialized in Electronic and Telecommunication Engineering from the University of Moratuwa, Sri Lanka, in 2017. Her research interests include Serverless Computing, Internet of Things (IoT), Distributed Systems and Reinforcement Learning. 
\end{IEEEbiography}
\begin{IEEEbiography}[{\includegraphics[width=1in,height=1.25in,clip,keepaspectratio]{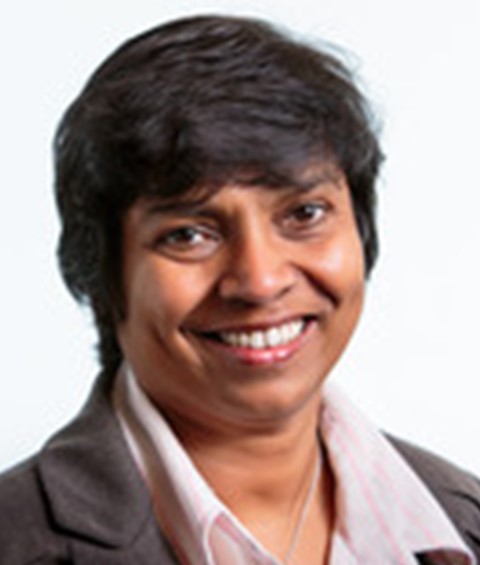}}]{Shanika Karunasekera}
is currently a Professor in the School of Computing and Information Systems and the Deputy Dean (Academic) in the Faculty of Engineeering and IT, University of Melbourne, Australia. She received her BSc degree in Electronic and Telecommunications Engineering from the University of Moratuwa, Sri Lanka, in 1990 and the PhD degree in electrical engineering from the University of Cambridge, U.K., in 1995. Her research interests include distributed computing, mobile computing, and social media analytics.
\end{IEEEbiography}
\begin{IEEEbiography}[{\includegraphics[width=1in,height=1.25in,clip,keepaspectratio]{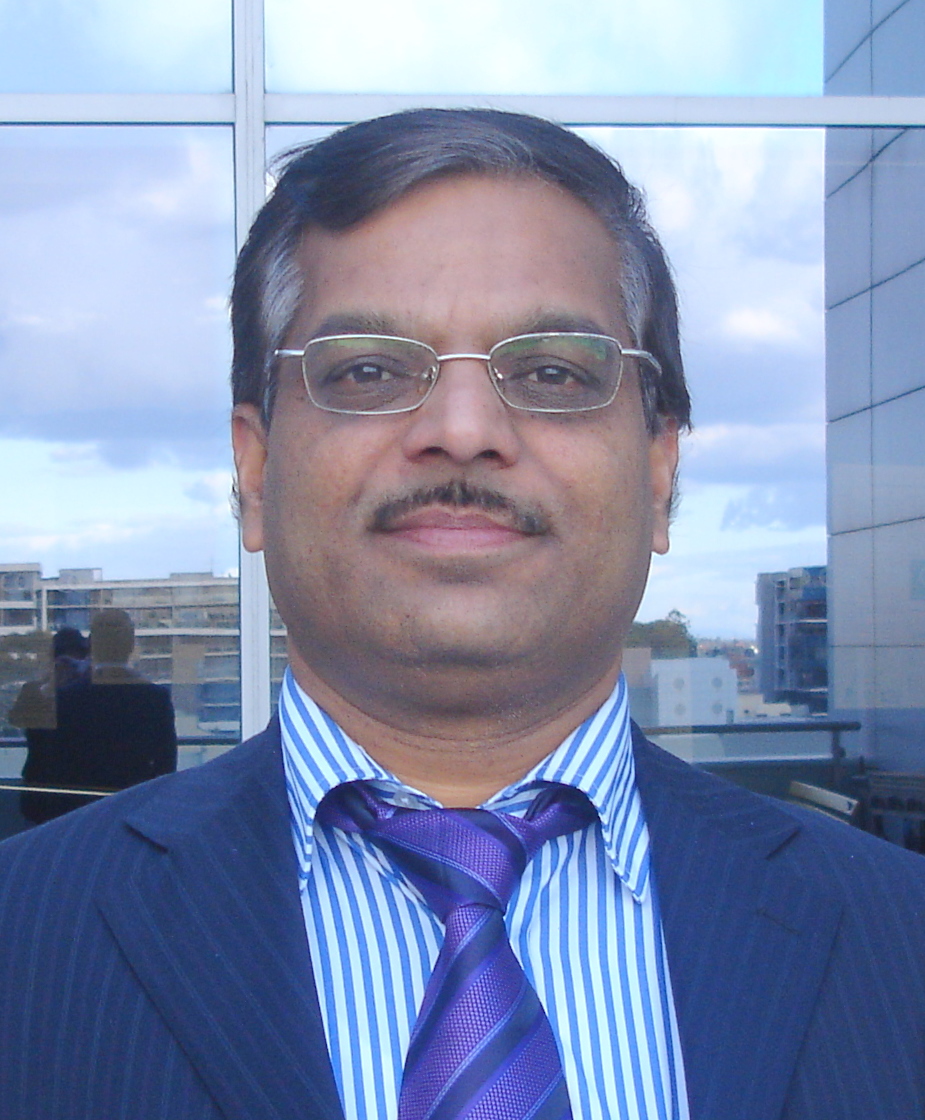}}]{Rajkumar Buyya}
is a Redmond Barry Distinguished Professor and Director of the Cloud Computing and Distributed Systems (CLOUDS) Laboratory at the University of Melbourne, Australia. He has authored over 625 publications and seven text books including "Mastering Cloud Computing" published by McGraw Hill, China Machine Press, and Morgan Kaufmann for Indian, Chinese and international markets respectively.  He is one of the highly cited authors in computer science and software engineering worldwide (h-index=160, g-index=345, 137100+ citations).
\end{IEEEbiography}

\vfill

\end{document}